%% file: intvarcat.tex
\documentclass[]{aa}

\usepackage{natbib}
\usepackage{graphicx}
\usepackage{amssymb}
\usepackage{amsmath}
\usepackage{longtable}
\usepackage{lscape}
\usepackage{afterpage}
\usepackage{txfonts}

\usepackage{color} 

\begin{document}

\date{\today}

\title{The catalog of variable sources detected by \emph{INTEGRAL} I: \\
Catalog and techniques \footnote{}}

\author{I.~Telezhinsky\inst{1,2}
	\and
	D.~Eckert\inst{3,4,5}
	\and
	V.~Savchenko\inst{4,5}
	\and
	A.~Neronov\inst{4,5}
	\and
	N.~Produit\inst{4,5}
	\and	
	T.J.-L.~Courvoisier\inst{4,5}
	} 
\authorrunning{Telezhinsky~I.~et~al.}
\titlerunning{The catalog of variable sources detected by \emph{INTEGRAL} I}
	\offprints{telezhinsky@gmail.com \\
* Table 3 is also available in electronic form at the CDS via anonymous ftp to cdsarc.u-strasbg.fr (130.79.128.5) or via http://cdsarc.u-strasbg.fr/viz-bin/qcat?J/A+A/vol/page}
\institute{
	DESY Zeuthen, Platanenallee 6, 15738 Zeuthen, Germany
	\and
	Astronomical Observatory of Kiev University, Observatorna 3, 04058 Kiev, Ukraine
	\and
	INAF/IASF Milano, Via E. Bassini 15, 20133 Milano, Italy
	\and
	ISDC Data Centre for Astrophysics, Chemin d'Ecogia 16, CH--1290 Versoix, Switzerland
    \and
	Observatoire Astronomique de l'Universit\'e de Gen\`eve, Chemin des Maillettes 51, CH--1290 Sauverny, Switzerland 
	}
\date{Received  / Accepted }

\abstract
{
During its 6 years of operation, \emph{INTEGRAL}/ISGRI has detected more than 500 sources. Many of these sources are variable. Taking into account that nearly half of \emph{INTEGRAL}/ISGRI sources are new and many of them remain unidentified, the variability properties of the sources can provide additional constraints to help us to classify and identify the unknown sources.
}
{
To study the variability properties of the sources detected by \emph{INTEGRAL}/ISGRI, we develop a method to quantify the variability of a source. We describe here our techniques and compile a catalog of the sources that fit our criteria of variability.
}
{
We use the natural time binning of \emph{INTEGRAL} observations called Science Window ($\approx 2000$ seconds) and test the hypothesis that the detected sources are constant using a $\chi^2$ all-sky map in three energy bands (20-40, 40-100, 100-200 keV). We calculate an intrinsic variance of the flux in individual pixels and use it to define the fractional variability of a source. The method is sensitive to the source variability on timescales of one Science Window and higher. We concentrate only on the sources that were already reported to be detected by \emph{INTEGRAL}.
}
{
We present a catalog of 202 sources found to be significantly variable. For the catalog sources, we give the measure of variability and fluxes with corresponding errors in the 20-40, 40-100 and 100-200 keV energy bands, and we present some statistics about the population of variable sources. The description of the physical properties of the variable sources will be given in a forthcoming paper.
}
{} 

\keywords{Gamma-rays: observations, catalogs}

\maketitle

\section{Introduction}

\emph{INTEGRAL} (INTErnational Gamma Ray Astrophysics Laboratory, \citet{Winetal03}) was launched in 2002 and since then has performed high-quality observations in the energy band from 3 keV up to $\sim 10$ MeV. The \emph{INTEGRAL} payload consists of two main soft gamma-ray instruments (the imager IBIS \citep{Ubeetal03}, and the spectrometer SPI \citep{Vedetal03}) and two monitors (in X-rays JEM-X \citep{Lunetal03}, and in optical OMC \citep{Masetal03}). The wide field-of-view of the imager IBIS provides an ideal opportunity to survey the sky in hard X-rays.\\

During its first 6 years in orbit, \emph{INTEGRAL} has covered nearly the whole sky. The observational data have been mainly used to study the soft gamma-ray emission from the Galactic plane (GP) \citep{Bouetal05, Krietal06} through the Galactic plane scans and the Galactic centre (GC) \citet{Beletal04, Revetal04, Beletal06, Krietal06} through the Galactic centre deep exposure programme. A number of papers have already presented general surveys \citep{Bazetal06, Biretal06} of the sky as well as of specific regions \citep{Gotetal05, Haretal06, Moletal04} and population types \citep{Baretal06, Basetal06, Sazetal07, Becetal06, Becetal09}.\\

The majority of the classified sources detected by \emph{INTEGRAL} are either low and high mass X-ray binaries (LMXBs and HMXBs) or AGNs \citep{Bodetal07}. However, a significant fraction of the detected sources remain unidentified. A special approach to population classification is required for the GC region to resolve the population types because of the high density of sources. Fortunately, the physics of the sources may help us to unveil their type. Indeed, the bulk of the \emph{INTEGRAL} sources are accreting systems that are expected to be intrinsically variable on multiple timescales depending on the source type and the nature of the variability. For instance, X-ray binaries (XRBs) may exhibit variability on timescales that range from milliseconds (supporting the idea that emission originates close to the compact object in the inner accretion radius) to hours and days, indicating that the variability can originate throughout the accretion flow at multiple radii and propagate inwards to modulate the central X-ray emission \citep{AreUtt05}. This idea is supported by the known correlation between millisecond/second and hour/day scale variability in XRBs \citep{Utt03}. LMXBs may exhibit flaring behavior with an increase in both emission intensity and hardness over a period of a few hundred to a few thousand seconds. X-ray bursts with rise times of a few seconds and decay times of hunderds of seconds or even several hours \citep{Baretal02} are also common to these objects. On the other hand, HMXBs are known to exhibit variability on timescales ranging from a fraction of a day up to several days, generated by the clumpiness of the stellar wind acreting onto the compact object \citep{Ducetal09}. Hour-long outbursts caused by variable accretion rates are observed in supergiant fast X-ray transients, a sub-class of HMXBs discovered by INTEGRAL \citep{Rometal2009}. Owing to their larger size, AGNs of different types exhibit day-to-month(s) variability depending on the black hole mass \citep{IshCou09}. Gamma-ray loud blazars have variability timescales in the range from $10^{1.6}$ to $10^{5.6}$~s \citep{LiaLiu03}. Therefore, a list of \emph{INTEGRAL} sources with quantitative measurements of their variability would be an important help to classifying the unidentified sources and more detailed studies of their physics.\\

The variability of \emph{INTEGRAL} sources was addressed in the latest 4th IBIS/ISGRI survey catalog paper \citep{Biretal09} when the authors performed the so-called \textit{bursticity} analysis intended to facilitate the detection of variable sources.\\

Here we present a catalog of \emph{INTEGRAL} variable sources identified in a large fraction of the archival public data. In addition to standard maps produced by the standard data analysis software, we compiled a $\chi^2$ all-sky map and applied the newly developed method to measure the fractional variability of the sources detected by the IBIS/ISGRI instrument onboard \emph{INTEGRAL}. The method is sensitive to variability on timescales longer than those of single ScW exposures ($\approx 2000$ seconds), i.e., to variability on timescales of hour(s)-day(s)-month(s). The catalog is compiled from the sources detected in the variability map. In addition, we implemented an online service providing the community with all-sky maps in the 20-40, 40-100, and 100-200 keV energy bands generated during the course of this research.\\

In the following, we describe the data selection procedure and the implemented data analysis pipeline (Sect.~\ref{sec:datana}). In Sect.~\ref{sec:method}, we outline our systematic approach to the detection of variability in \emph{INTEGRAL} sources and describe our detection procedure in Sect.~\ref{sec:detect}. We compile the variability catalog in Sect.~\ref{sec:catvar}. In Sect.~\ref{sec:skyview}, we briefly describe the implemented all-sky map online service. We make some concluding remarks in Sect.~\ref{sec:conclu}.

\section{Data and analysis}
\label{sec:datana}

\subsection{Data selection and filtering}

Since its launch, \emph{INTEGRAL} has performed over 800 revolutions each lasting for three days. We utilized the ISDC Data Centre for Astrophysics \citep{Couetal03} archive\footnote{http://isdc.unige.ch} to obtain all public data available up to June 2009 and the Offline Scientific Analysis (OSA) v. 7.0 to process the data. \\

\emph{INTEGRAL} data are organized into science windows (ScWs), each being an individual observation that can be either of pointing or slew type. Each observation (pointing type) lasts 1 -- 3 ksecs. For our analysis, we chose all pointing ScWs with an exposure time of at least 1 ksec. We filtered out revolutions up to and including 0025 belonging to the early performance verification phase, observations taken in staring mode, and ScWs marked as bad time intervals in instrument characteristics data including ScWs taken during solar flares and radiation belt passages. Finally, after the reconstruction of sky images we applied the following statistical filtering. We calculated the standard deviation of the pixel distribution for each ScW and found the mean value of standard deviations for the whole data set. We then rejected all the ScWs in which the standard deviation exceeded the mean for the whole data set by more than 3$\sigma$. We assumed the distribution of standard deviations of individual and independent ScWs to be normal. While calculating standard deviations in individual ScWs, image pixels were assumed to be independent. Thus, the filtering procedure allowed us to remove all ScWs affected by a high background level.  In the end, 43~724 unique pointing-type ScWs were selected for the analysis, giving us a total exposure time of 80.0~Msec and a more than 95 percent sky coverage.

\subsection{Instrument and background}

In the present study we use only the low-energy detector layer of the IBIS coded-mask instrument, called ISGRI (\emph{INTEGRAL} Soft Gamma Ray Imager, \citet{Lebetal03}), which consists of 16~384 independent CdTe pixels. It features an angular resolution of $12^{\prime}$ (FWHM) and a source location accuracy of $\sim$1 arcmin, depending on the signal significance \citep{Groetal03}. Its field of view (FOV) is $29^{\circ} \times 29^{\circ}$. The fully-coded part of the FOV (FCFOV), i.e., the area of the sky where the detector is fully illuminated by the hard X-ray sources, is $9^{\circ} \times 9^{\circ}$. It operates in the energy range between 15 keV and 1 MeV.\\

Over short timescales, the variability of the background of the instrument is assumed to be smaller than the statistical uncertainties. However, this is not the case for mosaic images constructed from long exposures. In general, it is assumed that the mean ISGRI background in each individual pixel changes very little with time, and therefore the standard OSA software provides only one background map for the entire mission. During the construction of the all-sky map, we noted that the quality of the mosaics of the extragalactic sky region depends on the time period over which the data were taken. We therefore, concluded that the long-term variation in the background of the instrument \citep{Lebetal05} significantly affects the extragalactic sky mosaic. On the other hand, in the GC and inner GP regions (l$\left\lbrace-90;90\right\rbrace$, b$\left\lbrace -20;20\right\rbrace$) the standard background maps provided by OSA provide better results (noise distributions are narrower). This might be because of the large number of bright sources and the Galactic ridge emission \citep{Krietal06}, although we leave this question open for the future research.\\

To produce time-dependent background maps, we extracted raw detector images for each \emph{INTEGRAL} revolution (3 days) and calculated the mean count rate in each individual pixel during the corresponding time period. To remove the influence of the bright sources on the neighboring background, we fitted and removed these sources from the raw detector images, i.e., in each ScW we constructed a model of the source pattern on the detector (pixel illumination fraction, PIF) and fitted the raw detector images using the model
\begin{equation}
S_{k,l}=\sum_{i=1}^M f_i \times PIF_{k,l}+B,
\end{equation}
where $S_{k,l}$ are the detector count rate, $PIF_{k,l}$ are the respective pattern model of source $i$ in the detector pixel with coordinates $(k,l)$, $f_i, i=1..M$ is the flux of source $i$ in the given ScW, and $B$ is the mean background level. This procedure was applied to all the detected sources in the FOV. The stability of the fitting procedure was tested using a large set ($>1~000$) of simulated ScWs with variable source fluxes. The results of the fit were normally distributed around the expected source flux, and therefore we can conclude that our procedure is sufficiently accurate to remove the point sources
from the construction of the background maps. The results of the fitting procedure were then used to create a transformed detector image, $\hat S_{k,l}$, defined as
\begin{equation}
\hat S_{k,l}=S_{k,l}-\sum_{i=1}^M f_i \times PIF_{k,l}.
\end{equation}
Background maps were then constructed by averaging the transformed detector images of a given data set.

From our time-dependent background maps, we found that the shape of the ISGRI background varies with time, in particular after each solar flare. A long-term change in the background was noticed as well. This result agrees with the findings of \citet{Lebetal05}. To take these variations into account, we generated background maps for each spacecraft revolution and in the image reconstruction step applied them to the extragalactic sky region.

Besides the real physical background of the sky, there is also artificial component, because IBIS/ISGRI is a coded-mask instrument with a periodic mask pattern. Therefore, the deconvolution of ISGRI images creates structures of fake sources that usually appear around bright sources. Apart from the periodicity of the mask, insufficient knowledge of the response function leads to residuals in the deconvolved sky images. The orientation of the spacecraft changes from one observation of the real source to another, so fake sources and structures around the real source contribute to the noise level of the local background. To reduce this contribution, we used a method described in Sect.~\ref{subsec:imarec}.\\

\subsection{Image reconstruction}
\label{subsec:imarec}
\begin{figure*}[!th]
\begin{center}
\includegraphics[width=0.95\textwidth]{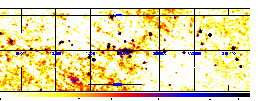}
\includegraphics[width=0.95\textwidth]{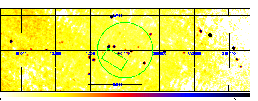}
\caption{Inner parts (36$^{\circ}$ by 12$^{\circ}$) of the INTEGRAL/ISGRI all-sky maps in Galactic coordinates, Aitoff projection. The significance image (top) in the 20-40 keV energy band, has square root scaling. The bottom image shows the corresponding intrinsic variance map, and also has square root scaling. The circle shows the inner 4$^{\circ}$ for which the variability background extraction was made from the box region.}
\label{fig:gc}
\end{center}
\end{figure*}

After producing the background maps as described in the previous subsection, we started the analysis of the data using the standard Offline Scientific Analysis (OSA) package, version 7.0, distributed by ISDC \citep{Couetal03}. For image reconstruction, we used a modified version of the method described in \citet{Ecketal08}.  It is known that screws and glue strips attaching the IBIS mask to the supporting structure can introduce systematic effects in the presence of very bright sources \citep{Nevetal09}. To remove these effects, we identified the mask areas where screws and glue absorb incoming photons, and we disregarded the pixels illuminated by these mask areas for the 11 brightest sources in the hard X-ray band. No more than 1\% of the detector area was disregarded for each of the brightest sources. For weaker sources, the level of systematic errors produced by the standard OSA software was found to be consistent with the noise, so the modified method was not required. Finally, we summed all the processed images weighting by variances to create the all-sky mosaic. For this work, we produced mosaics in 3 energy bands (20-40, 40-100, and 100-200 keV). Both our all-sky map images and corresponding exposure maps are available online and we direct the reader to our online web service\footnote{http://skyview.virgo.org.ua}. As an example, we provide here the image of the inner part (36$^{\circ}$ by 12$^{\circ}$) of the Galaxy in the 20-40 keV energy band (see Fig.~\ref{fig:gc}).

\begin{figure*}[!t]
\begin{center}
\includegraphics[width=0.40\textwidth]{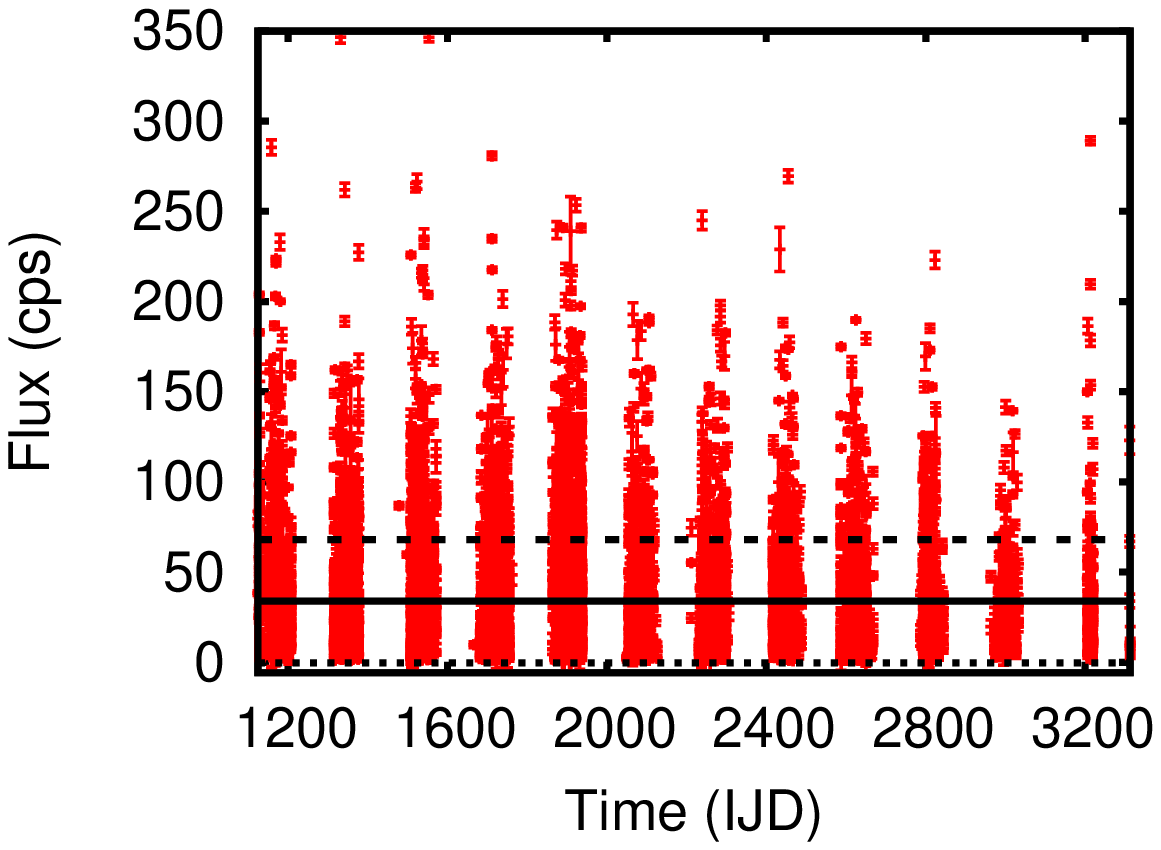}
\includegraphics[width=0.14\textwidth]{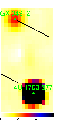}
\includegraphics[width=0.40\textwidth]{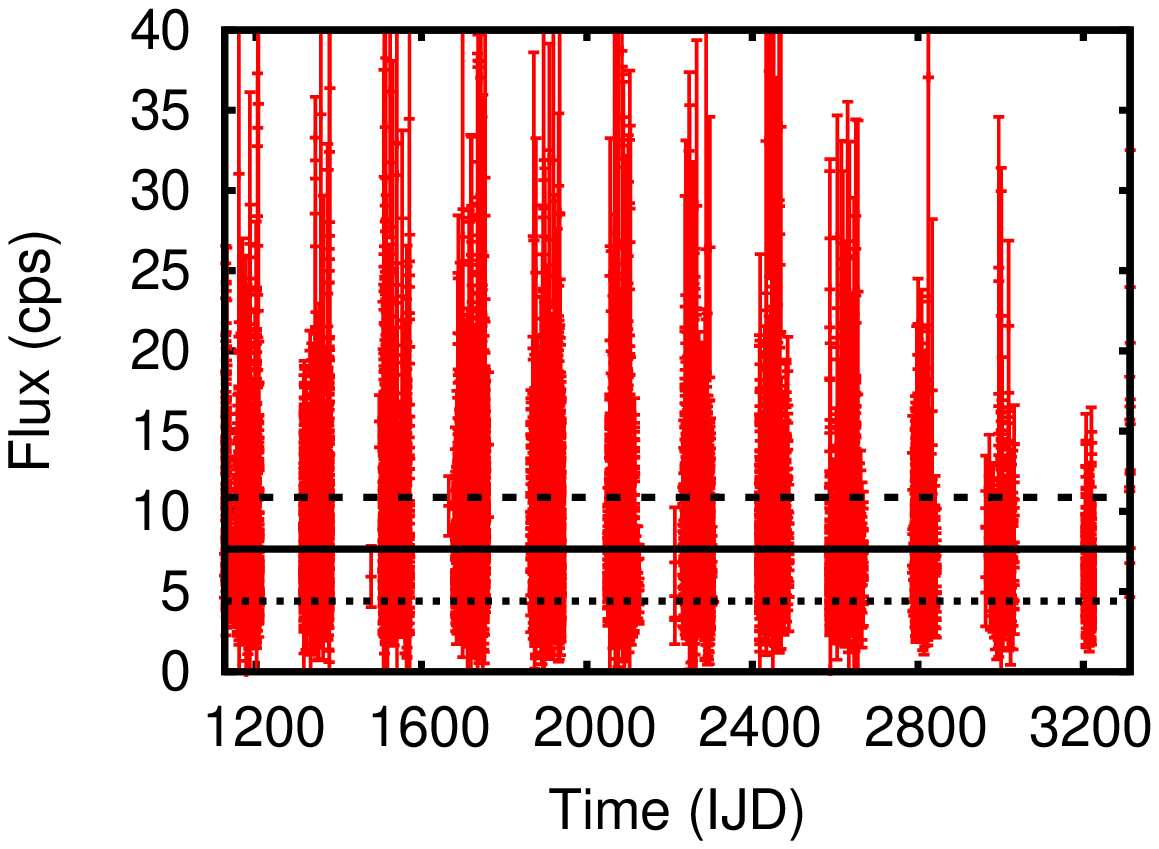}
\caption{Lightcurves and variability map of HMXB 4U~1700-377 and LMXB GX~349+2. The solid line indicates the mean flux of the sources during the observation time, the dotted line shows the mean flux minus $S_{int}$, the dashed line shows the mean flux plus $S_{int}$.}
\label{fig:lc}
\end{center}
\end{figure*}

\section{Method of variability detection}
\label{sec:method}
The variability of \emph{INTEGRAL} sources can be analyzed in a standard way by studying the inconsistency of the detected signal with that expected from a constant source by performing the $\chi^2$ test. Here we consider introducing a variability measurement for the \emph{INTEGRAL} sources and show how to apply it to the specific case of the coded-mask instrument. For an alternative approach based on the maximum likelihood function for the determination of intrinsic variability of X-ray sources the reader is referred to \citet{Almetal00} and \citet{Becetal07}.

The \emph{INTEGRAL} data are naturally organized by pointings (ScW) with average duration of $\sim 1-3$~ksec. Therefore, the simplest way to detect the variability of a source on ksec and longer timescales is to analyse the evolution of the flux from the source on a ScW-by-ScW basis. We define $F_i$ and $\sigma_i^2$ to be the flux and the variance of a given source, respectively, in the $i$-th ScW. The weighted mean flux from the source is then given by
\begin{equation}
\langle F \rangle=\frac {\sum_{i=1}^N \frac {F_i}{\sigma_i^2}} {\sum_{i=1}^N \frac{1}{\sigma_i^2}},
\end{equation}
where $N$ is the total number of ScWs. The variance of the source's flux, which is the mean squared deviation of the flux from its mean value during the observation time, is given by
\begin{equation}
S_{tot}^2 = \frac {\sum_{i=1}^N \frac{(F_i - \langle F \rangle)^2}{\sigma_i^2}} {\sum_{i=1}^N \frac{1}{\sigma_i^2}} = \chi^2\sigma^2,
\label{SV}
\end{equation}
where $\chi^2 = \sum_{i=1}^N \frac{(F_i - \langle F \rangle)^2}{\sigma_i^2}$ and $\sigma^2 = \left(\sum_{i=1}^N \frac{1}{\sigma_i^2}\right)^{-1}$ is the variance of the weighted mean flux.

However, in addition to intrinsic variance of the source, this value includes the uncertainty in the flux measurements during individual ScWs, i.e., the contribution of the noise. If the source variance is caused only by the noise, i.e., $F_i = \langle F \rangle \pm \sigma_i$, Eq.~(\ref{SV}) is given by $S_{noise}^2 = N \sigma^2$. To eliminate the noise contribution, we can subtract the noise term of the variance from the source variance and derive the \emph{intrinsic variance} of the source

\begin{equation}
S_{int}^2 = \chi^2\sigma^2 - N\sigma^2.
\label{intvar}
\end{equation}
When all measurement errors are equal ($\sigma_i = \sigma_0$, $\sigma^2 = \sigma_{0}^2/N$), our case reduces to the method used by \cite{Nanetal97}

\begin{equation}
S_{int}^2 = \frac {1}{N} \sum_{i=1}^N (F_i - \overline{F})^2 - \sigma_{0}^2,
\end{equation}
where $\overline{F}$ is the unweighted mean flux and $S_{int}^2$ is called the \emph{excess variance}. In the absence of measurement errors, our case reduces to the standard definition of the variance

\begin{equation}
S_{int}^2 = \frac {1}{N} \sum_{i=1}^N (F_i - \overline{F})^2.
\end{equation}

Given that different sources have different fluxes, the variability of sources can be quantified by using the normalized measure of variability, which we call here the \emph{fractional variability}
\begin{equation}
V = \frac {S_{int}}{\langle F \rangle}.
\label{simplefracvar}
\end{equation}
However, in reality, if one were to apply the above method to detect the variable sources in a crowded field (i.e., containing many sources) of a coded-mask instrument such as IBIS, one would infer {\it all} the detected sources to be highly variable. This is because in coded-mask instruments, each source casts a shadow of the mask on the detector plane. If there are several sources in the field of view, each of them produces a shadow that is spread over the whole detector plane. Some detector pixels are illuminated by more than one source. If the signal in a detector pixel is variable, one can tell, only with a certain probability, which of the sources illuminating this pixel is responsible for the variable signal. Thus, in a coded-mask instrument, the presence of bright variable sources in the field of view introduces an ``artificial'' variability for all the other sources illuminating the same pixels. Since the overlap between the PIF of the bright variable source and the sources at different positions on the sky varies with the position on the sky, one is also unable to determine in advance the level of this ``artificial'' variability in a given region of the deconvolved sky image.

To overcome this difficulty, one has to measure the variability of the flux not only directly in the sky pixels at the position of the source of interest, but also in the background pixels around the source. Obviously, the ``artificial'' variability introduced by the nearby bright sources is similar in the adjacent background pixels to that in the pixel(s) at the source position. Therefore, one can produce the variability map for the whole sky and compare the values of variability at the position of the source of interest to the mean values of variability in the adjacent background pixels. The variable sources should be visible as local excesses in the variability map of the region of interest. If a source can be localized in the variability image, then the true fractional variability of the source is calculated as

\begin{equation}
V_r = \frac {\sqrt{S_{int,s}^2 - S_{int,b}^2}} {\langle F_s \rangle - \langle F_b \rangle},
\label{fracvar}
\end{equation}
where the subscript $b$ represents the values of the background in the area adjacent to the source and the subscript $s$ the values taken from the source position.\\

To illustrate the method, we present the lightcurves (Fig.~\ref{fig:lc}) of two objects that are typical bright \emph{INTEGRAL} sources: the HMXB 4U~1700-377, which is a very bright and very variable source ($V_{r} \simeq 104$~\%), and the LMXB GX~349+2, which is a moderately bright and variable source ($V_{r} \simeq 45$~\%). The solid line indicates the mean flux of the sources, $\langle F \rangle$. We can see that the mean flux deviation (dotted lines), calculated as the square root of the intrinsic variance, $S_{int}^2$, measures the average flux variation of the sources during the corresponding time. However, we note that in the present case we consider calculations based solely on a lightcurve. If one wishes to obtain a fractional variability value dividing the mean flux deviation by the mean flux, one will obtain the $V$ value, but not $V_{r}$, i.e., the contribution of bright variable neighbor sources is not treated properly. It is impossible to extract the variability of the background, $S_{int,b}$, and the mean background flux $\left\langle F_b \right\rangle$ using the source lightcurve only. A number of lightcurves of the neighboring pixels should also be compiled to estimate $S_{int,b}$ and $\left\langle F_b \right\rangle$. This number should be sufficiently high to obtain good estimates. Therefore, an all-sky approach is justified. In the current example, no sources are much brighter in the vicinity of the ones considered that could strongly affect them, so the difference between $V$ and the catalog $V_r$ value is around 3\%, but for the weak sources in the vicinity of bright ones the difference would be higher.

Looking at Eq.~\ref{fracvar}, one can indeed see that the effect of ``artificial'' fractional variability is strong for moderate and faint sources in the vicinity of the bright variable sources, while for bright sources the effect is small. The ``artificial'' variability introduced by the bright sources in their vicinity ($S_{int,b}^2$ for the surrounding sources) is in range from a fraction of a percent up to a few percent of their own variability (dictated by the PIF accuracy). When we consider the persistent source in the vicinity of the bright variable source, $S_{int,s}^2$ is defined by $S_{int,b}^2$ only (i.e., by variability introduced by the bright variable source). For moderate or faint sources, $S_{int,b}$ may well be comparable to their own flux, and if we apply Eq.~\ref{simplefracvar} directly we will infer substantial fractional variability, which may well be between ten and fifty percent, or even higher. The bright sources are less sensitive to this effect because $S_{int,b}$ will be only a small fraction of their flux. We checked these conclusions by performing simulations of a moderate persistent source ($F = 1$~ct/s) in the vicinity of a bright variable source ($\langle F \rangle = 20$~ct/s). By applying Eq.~\ref{simplefracvar} directly to measure the fractional variability of a moderate source, we found that $V \simeq 25$\% while Eq.~\ref{fracvar} infered that the source was constant, i.e., $V_r \simeq 0$\%. 

\begin{figure*}[!th]
\begin{center}
\includegraphics[width=0.61\textwidth]{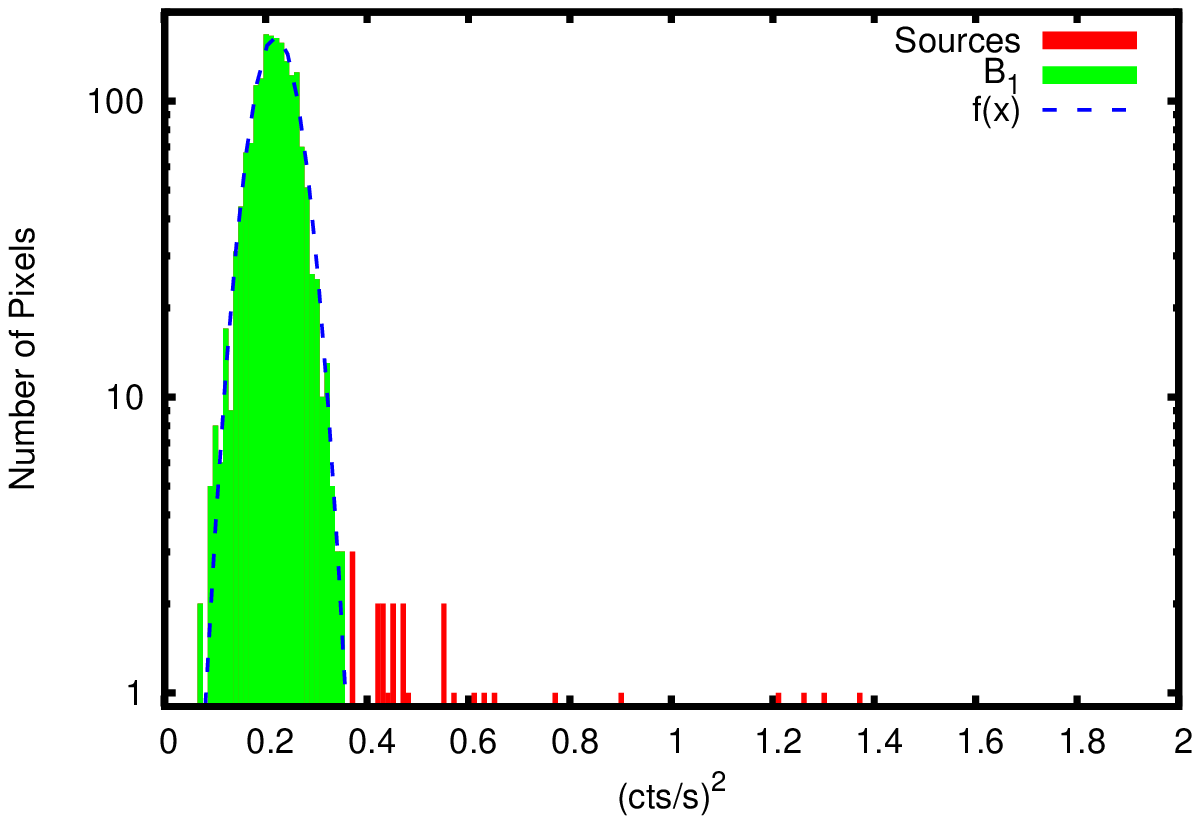}
\includegraphics[width=0.37\textwidth]{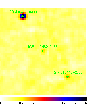}
\includegraphics[width=0.61\textwidth]{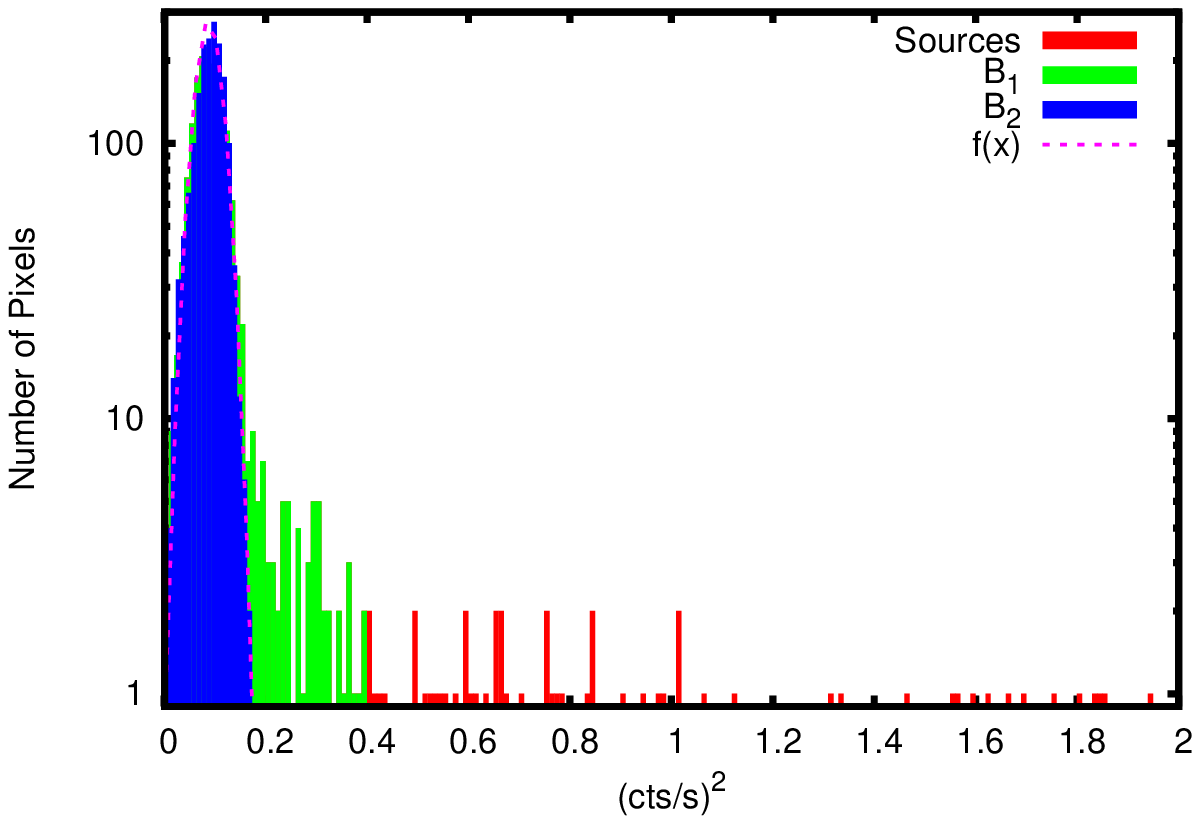}
\includegraphics[width=0.37\textwidth]{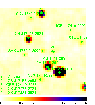}
\caption{Top: the distributions of the intrinsic variability in pixels of $3.5^{\circ} \times 3.5^{\circ}$ area (shown nearby) centered on the IGR~J18450-0435. In green is $B_1$ distribution in the range $(min, 2m-min)$ here representing the local intrinsic variance background, in red is the sources contribution, f(x) is the gaussian distribution. Bottom: the $3.5^{\circ} \times 3.5^{\circ}$ area (shown nearby) centered on the GC source 1A~1743-288. In green is $B_1$ distribution in the range $(min, 2m-min)$, in blue is $B_2$ distribution from the empty region in GC here representing the local intrinsic variance background, in red and f(x) are same as above.}
\label{fig:sintbhist}
\end{center}
\end{figure*}

In the course of simulations, we also checked the behavior of the ``artificial'' variability in the case a moderate persistent source situated at the ghost position of the bright variable source. We considered two situations: a) when the mosaic image consisted mostly of ScWs in a configuration being determined almost entirely by the spacecraft orientation, which remained constant (i.e., sky region of the Crab), and b) when the mosaic image contained only a chance fraction of ScWs in that specific configuration because of different spacecraft orientations (i.e., sky region of the Cyg X-1). The simulations showed that the flux and therefore the variability measurement of the mosaic deviated from the input source parameters in case a) only, while in case b), there were no significant deviations. As expected in case a), the moderate source was affected. This was caused by the coincidence of the sources shadowgrams. The deconvolution procedure was unable to distinguish the sources correctly on the ScW level, therefore the mosaic results were affected. We detected an incorrect flux and variability for the moderate source. In reality, this particular simulated situation is very rare (see Sect.~\ref{sec:catvar} for discussion of this situation in real case). Even if the constant orientation of some observation is kept, different observation patterns applied during the observation will significantly reduce the undesirable effect considered in situation a). 
\\

\section{The detection of variable sources}
\label{sec:detect}

For our study, we used the latest distributed \emph{INTEGRAL} reference catalog\footnote{http://isdc.unige.ch/?Data+catalogs} \citep{Ebietal03} version 30 and selected the sources with ISGRI\_FLAG == 1, i.e., all the sources ever detected by IBIS/ISGRI.\\

Based on the aforementioned method, we compiled the instrinsic variance maps ($S_{int}^2$) of the \emph{INTEGRAL} sky in three energy bands (20-40, 40-100, and 100-200 keV). This was accomplished by performing pixel operations following Eq.~(\ref{intvar}) on the constructed all-sky mosaic maps of $\chi^2$, $\sigma^2$ (variance), and $N$, which is the map showing the number of ScWs used in a given pixel for the all-sky mosaic. As an example, the instrinsic variance image of the inner region of our Galaxy is given in Fig.~\ref{fig:gc} (see our online service for all-sky maps).\\

After compiling an intrinsic variance map in each band, we calculated the local (or background) intrinsic variance, $S_{int,b}^2$, and its scatter, $\Sigma$, in the region around each catalog source. This was performed by the following scheme. First, the values of the mean $m$, the minimum $min$, and their difference were calculated in a square of $3.5^{\circ} \times 3.5^{\circ}$ centered on the source position. We then assumed that the pixel values in the corresponding area are distributed normally and there is the always-positive contribution from the sources in the field. Since the sources occupy a small fraction of the considered sky region, the initial mean value, $m$, is almost unaffected by their presence because of their small contribution. To reject the source contribution and to obtain the parameters of the normal component, we calculated the mean and the standard deviation in the range $(min, 2m-min)$. The newly found mean value is $S_{int,b}^2$ in Eq.~\ref{fracvar} and the standard deviation shows its scatter $\Sigma$. The detectability of the sources in the intrinsic variance map is then defined by the condition that $S_{int}^2 \geq S_{int,b}^2 + 3 \Sigma$. For an illustration (see top of Fig.~\ref{fig:sintbhist}) we present a region around INTEGRAL source IGR~J18450-0435 with respective distributions. The green solid area is the distribution in the range $(min, 2m-min)$ with the mean value representing $S_{int,b}^2$ for the current source, and in red the always-positive contribution from the sources in the field is given. The distribution in the range $(min, 2m-min)$ is well fitted by the Gaussian shown with a dashed line. The top of Fig.~\ref{fig:sintbhist} justifies well the approximate rejection of the source contribution to the overall pixel value distribution in the field. Applying this rejection procedure allows us to obtain the true scatter in the background rather than the scatter in the overall distribution (including sources), which is obviously higher.

The detection of variable sources in the innermost region of our Galaxy (circle of 4$^\circ$ from the GC, see Fig.~\ref{fig:gc}) was performed differently because it is a crowded field and therefore a large number of sources contribute to the intrinsic variance background of each other. In contrast to the individual source case, the sources in the inner GC region occupy a significant fraction of the region around the source of interest and therefore influence the $m$ value significantly. This results in improper estimation of the background distribution if one applies the rejection procedure based on $(min, 2m-min)$ range ($B_1$ distribution at the bottom of Fig.~\ref{fig:sintbhist}). In place of calculating the intrinsic variance background and its scatter for each GC source individually, these values were therefore estimated from an empty field near the GC (box at the bottom image of Fig.~\ref{fig:gc}) and assumed to be equal for all the sources in the GC region. The $B_2$ distribution shown at Fig.~\ref{fig:sintbhist} (bottom) is accurately determined and well fitted by the Gaussian shown by the dashed line.\\

\section{The catalog of variable sources}
\label{sec:catvar}

The search for variable sources from the reference catalog was performed on the maps generated from ScWs divided into three equal subsequent time periods (maps 1,2, and 3, approximately 2 years each). This was done to detect transient sources that would be difficult (or even impossible) to detect on the map integrated over the whole time period (map T). The search was performed separately on each map (1,2,3) and in each energy band. The results of the search were put into one list from which the sources were selected. Finally, the search was performed on the total map (T) to find sources that were detected only on the map integrated over the whole time period and identify the sources that were active only during specific time periods. The resulting catalog of variable sources detected by \emph{INTEGRAL} can be found in Table~\ref{varcat}.

\begin{figure}[!t]
\begin{center}
\includegraphics[width=0.99\columnwidth]{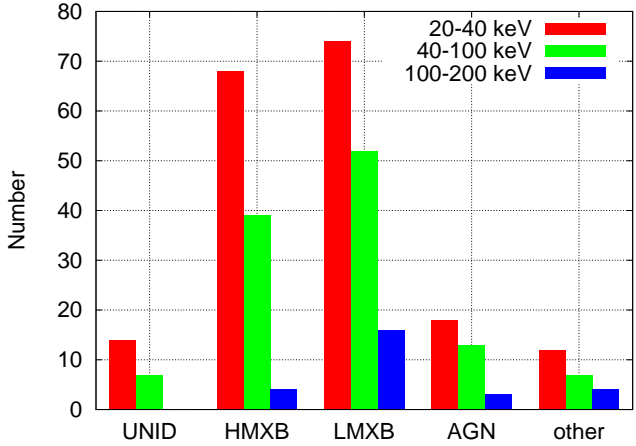}
\caption{Number of significantly variable sources detected in different energy bands classified by types.}
\label{chart:poptypes}
\end{center}
\end{figure}

Figure~\ref{chart:poptypes} shows the number of significantly variable sources from our catalog for each source type (HMXB, LMXB, AGN, unidentified, and miscellaneous). The majority of the variable sources ($\sim$76\%) in all energy bands are Galactic X-ray binaries. We see that in the 100-200 keV band there are four times more LMXBs than HMXBs. The remaining variable sources are AGNs ($\sim$10\%), unidentified ($\sim$7.5\%),  and other ($\sim$6.5\%) source types (cataclysmic variables, gamma-ray bursts, and pulsars). The number of significantly variable sources decreases with energy for each population type, which only reflects the sensitivity of the instruments.

\begin{figure}[!t]
\begin{center}
\includegraphics[width=0.99\columnwidth]{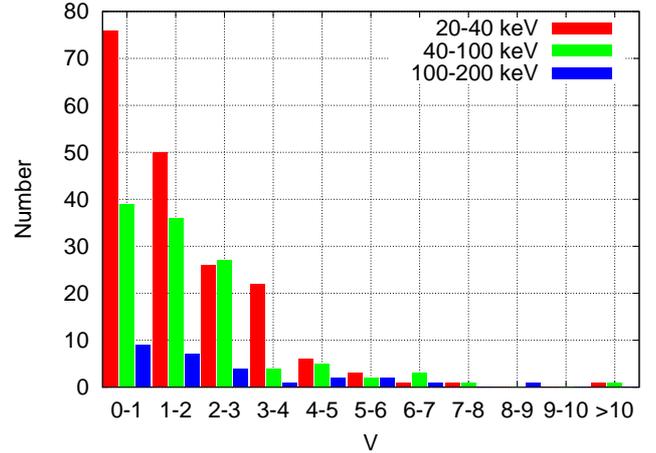}
\caption{Number of sources versus fractional variability of the sources detected in different energy bands.}
\label{chart:histo}
\end{center}
\end{figure}

\begin{figure}[!t]
\begin{center}
\includegraphics[width=0.99\columnwidth]{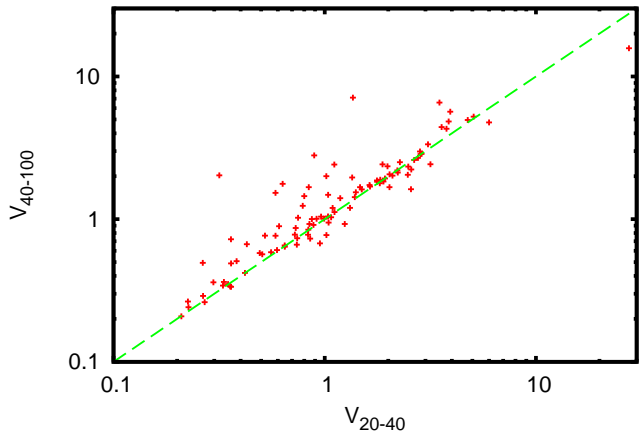}
\caption{Fractional variability of sources in the 20-40 keV band versus fractional variability in the 40-100 keV band.}
\label{fig:V1-V2}
\end{center}
\end{figure}

The distribution of the variability of sources from Table~\ref{varcat} is presented in Fig.~\ref{chart:histo}. The variability distribution is approximately normal with one evident outlier, the gamma-ray burst IGR J00245+6251 \citep{Vesetal05}. However, this is mainly caused by the upper limit to the mean flux of this source being too low. The gamma-ray burst IGR J00245+6251 is detected in all three energy bands. Figure~\ref{fig:V1-V2} shows the fractional variability in the 40-100 keV band versus the variability in the 20-40 keV band. The majority of the sources that are found to be variable in both the 20-40 keV and 40-100 keV energy bands exhibit nearly equal variability in both bands. 

\begin{figure}[!t]
\begin{center}
\includegraphics[width=0.99\columnwidth]{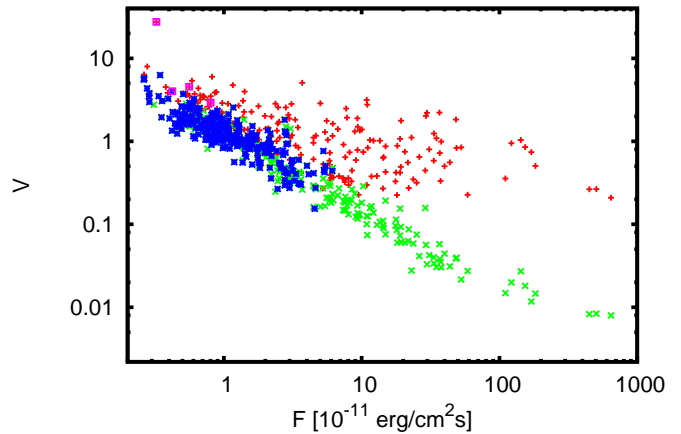}
\caption{Fractional variability (V) versus flux (F) for all significantly variable sources (red crosses) from Table~\ref{varcat}. Energy band is 20-40 keV. For comparison, the green crosses show the fractional variability detection threshold (3-$\sigma$) versus flux for all reference catalog sources detected in the significance map. The sources not detected at the variability map are indicated by blue stars and are coincident with their green cross counterparts. Pink squares indicate 3-$\sigma$ lower limits to the fractional variability of the sources that are not detected in the significance map.}
\label{fig:logV-logF}
\end{center}
\end{figure}

To show the detection threshold for the variability of a source, we compiled a diagram (see Fig.~\ref{fig:logV-logF}) where we plot the fractional variability versus flux for all detected variable sources along with the upper limit to the fractional variability versus flux for all the reference catalog sources detected in our significance map in 20-40 keV energy band. The upper limit was determined by substituting $S_{int,s}^2$ with $S_{int,b}^2 + 3 \Sigma$ in Eq.~(\ref{fracvar}). Although we chose all the sources detected in our significance map, we used the mean flux of the source because, unlike significance, it is an exposure-independent value. According to our definition, the fractional variability is also an exposure-independent value, so we plot it versus the exposure-independent mean flux to clearly see the detection threshold. We can see that starting from a limiting flux ($6.2\times10^{-11}$~ergs/cm$^2$s or $\sim10$~mCrab) nearly all catalog sources are found to be variable. The majority of the bright sources are binary systems, which explains why nearly all of them are found to be variable.

\begin{table*}[!h]
\begin{center}
\caption{The transient sources detected at the intrinsic variance map and not detected at the significance map. }
\begin{tabular}{|l|l|l|l|l|l|}
\hline
Name & Class & $V_{r,notnorm} \pm Err$ (c/s) & Exposure (ksec) & Map & Band\\
\hline
\input{varnstab.tex}
\hline
\end{tabular}
\label{transou}
\end{center}
\end{table*}

We also found a number of variable sources that have no counterparts in the significance map, which means that we were unable to measure the mean flux of these sources as it is compatible with the background mean flux. Hence, we are not able to give their fractional variability value, which is a normalized value and therefore tends to infinity with infinite errors. For these sources, we provide a 3-$\sigma$ lower limit to their fractional variability by taking a 3-$\sigma$ upper limit on their mean flux. Most of the sources are transient, and sometimes (in a specific observation period in a given energy band) the source is not detected in the significance map because the sensitivity of the instrument decreases with energy. Therefore, we can see that the variability map provides a tool to detect transient or faint but variable sources that would be missed in mosaics averaged over long timescales. To illustrate the detectability of the sources in the variability map, we provide a list (see Table~\ref{transou}) of the sources that are smeared out in the significance map because of their high exposures. The values given in the table are not normalized variability values, $V_{r,notnorm} = \sqrt{S_{int,s}^2 - S_{int,b}^2}$ along with their 3-$\sigma$ errors, $Err$. To verify that the sources that are absent in the significance maps but detected in the variability maps are not the result of the low detection threshold, we ran the same detection procedures on the mock catalog of 2500 false sources distributed randomly and uniformly over the sky. The test detected 11 of 2500 false sources seen in variability maps and absent from the significance maps, compared to 8 of 576 in the case of the real catalog. This means that our detection criteria are rather strict.

We comment on the inclusion of Crab in our catalog. It is known to be a constant source that is commonly used as a ``standard candle'' in high-energy astrophysics. There are two reasons why it appears in the catalog. The first is the deterioration of the detector electronics onboard the spacecraft. The loss of detector gain is around 10\% over the entire mission. Although this loss is partially compensated by the software, it introduces around 3-5\% variability in our method. The remaining variability can be ascribed to systematic errors in the spacecraft alignment \citep{Waletal03}, which for OSA 7.0 are of the order of 7 arcsec \citep{Ecketal09}, hence slightly different positions of the Crab are found in each observation. Since Crab is a very bright source, its Gaussian profile is very steep. When the peak is slightly offset, we measure a sharp decrease in the flux at the catalog position of Crab. The combination of the two effects leads to an artifical variability of Crab in all energy bands. A similar effect occurs in the pixels adjacent to the catalog position of Crab.
When the peak of the PSF is found at a slightly displaced position, we find a sharp increase in the flux in this pixel. Our interpretation is confirmed by the image of Crab in the variability map, where the closest pixels to the source are found to be very variable, creating a ring-like structure. Moreover, it has been found that the observed position of Crab does not coincide with the position of the pulsar \citep{Ecketal09}, which thus explains such an effect at the expected source position. To determine the influence of this effect on other sources, we inspected the 11 brightest sources in the 20-40 keV band and looked for a similar situation. In the case of Cyg~X-1 and Sco~X-1, the same effect, albeit weaker, was also seen. However, the derived value of their variability was not found to be affected by this effect. This effect contributes mainly to the variability of the pixels around the catalog position of these sources. Cyg X-1 and Sco X-1 are intrinsically very variable so the value extracted from the source position is much higher than for surrounding pixels, which is the opposite of the situation found in the Crab. For all the other sources, the influence of the misalignment was found to be negligible.

\begin{table*}[!h]
\begin{center}
\caption{Sources with additional error induced by the ``ghosts''.}
\begin{scriptsize}
\begin{tabular}{|l|l|l|l|l|l|l|l|l|l|l|l|l|l|l|l|}

\hline
\multicolumn{1}{|l|}{Name} & \multicolumn{5}{c|}{20-40 keV} &\multicolumn{5}{c|}{40-100 keV} &\multicolumn{5}{c|}{100-200 keV} \\
\cline{2-16}
\multicolumn{1}{|l|}{} & $V_r$ & $V_{r,err}$ & $G_{err}$ & Gexp & Exp & $V_r$ & $V_{r,err}$ & $G_{err}$ & Gexp & Exp & $V_r$ & $V_{r,err}$ & $G_{err}$ & Gexp & Exp \\

\hline
\input{ghostab.tex}
\hline
\end{tabular}
\label{ghostab}
\end{scriptsize}
\end{center}
\hspace{-6mm} Here, $V_r$ is the true fractional variability of the source, $V_{r,err}$ is the fractional variability error same as in catalog, $G_{err}$ is the error induced by the source, Gexp, in ksec, is the exposure time the source was affected by the ``ghost'', Exp, in ksec, is the total exposure time.
\end{table*}

Finally, we performed a test to find cases in which a source is coincident with the ghost of another source within one ScW (a case described in Sect.~\ref{sec:method}). We considered all the reference catalog sources and searched for ``ghost-source'' pairs in individual ScWs from the list used for our all-sky maps. If one of the sources in a given pair was present in our variability catalog, the pair was selected for further analysis. According to our simulations, if two sources are in the ghosts of each other, the fainter one loses up to half of its flux to the flux of the brighter one. If one of the sources is substantially brighter than the other, the relative distortion of the flux of the bright source is minor. Therefore, the flux of the source is significantly distorted if its ``ghost companion'' is brighter or comparable in brightness to the source itself. In the latter case, the ``ghost companion'' is also affected. Thus, we assume that if the source in the ghost is more than ten times fainter than its ``ghost companion'', its contaminating influence is negligible, whereas its own flux is contaminated significantly. After adopting this condition, we obtained a list of the sources affected by such position coincidences and the exposure times for each of them during which they where affected. For nearly all of the sources from the list, the fraction of exposures with distorted flux is less than 1\%, which infers a relative uncertainty in the average flux and fractional variability of the same order. This is much smaller than the error set on variability in our catalog and, as can be seen from the Fig.~\ref{fig:logV-logF} is smaller than the variability detection threshold even for the brighest sources. Nonetheless, a number of sources have larger than 1\% errors induced by the ghosts, which we indicate with a $^{g}$ superscript in the catalog and provide a Table~\ref{ghostab} where both errors are given. As can be seen, in all cases the ghost induced error is much smaller than the catalog variability error.

\section{The All-Sky online}
\label{sec:skyview}

To make our results available to the community, we decided to take advantage of the SkyView interface \citep{Mcgetal98} (i.e., a Virtual Observatory service available online\footnote{http://skyview.gsfc.nasa.gov} developed at and hosted by the High Energy Astrophysics Science Archive Research Centre (HEASARC)). SkyView provides a straightforward interface where users can retrieve images of the sky at all wavelengths from radio to gamma-rays \citep{Mcgetal98}. SkyView uses NED and SIMBAD name resolvers to translate names into positions and is connected to the HEASARC catalog services. The user can retrieve images in various coordinate systems, projections, scalings, and sizes as ordinary FITS as well as standard JPEG and GIF file formats. The output image is centered on the queried object or sky position. SkyView is also available as a standalone Java application \citep{Mcgetal97}. The ease-of-use of this system allowed us to incorporate INTEGRAL/ISGRI variability and significance all-sky maps into the SkyView interface. We developed a simple web interface for the SkyView Java application and have made all-sky mosaics available online.

\section{Concluding remarks}
\label{sec:conclu}

Our study of variability of the \emph{INTEGRAL} sky has found that 202 sources exhibit significantly variable hard X-ray emission. To compile the catalog of variable sources, we have developed and implemented a method to detect variable sources, and compiled all-sky variability maps. A search for variable sources from the \emph{INTEGRAL} reference catalog was carried out in three energy bands: 20-40, 40-100, and 100-200 keV. The variable sources were detected in all studied energy bands, although their number sharply decreases with increasing energy. A number of sources were detected only during specific time periods and not detected on the map integrated over the whole time period. These sources are most likely transient. On the other hand, some sources were found to be variable only on the total variability map. This means that they might be persistent and not extremely variable sources.

We found that around 76\% of all variable sources of our catalog are binary systems and nearly 24\% of variable sources are either AGNs, unidentified, or other source types. The variability measurements of our catalog sources have rather normal distributions in all energy bands. We found that in the majority of cases the variability of the source in the first band correlates with its variability in the second band. We derived the limits to the fractional variability value to be detected as a function of the source flux (Fig.~\ref{fig:logV-logF}). We also found that variability map can be a tool to detecting transient or faint but variable sources that would be missed in mosaics averaged over long timescales.  In a forthcoming paper, we will discuss in more detail the properties of the variable sources detected during this study in order to gain some physical insights into the population of hard X-ray sources.\\

Finally, we emphasize that the sky maps generated during this study represent 6 years of \emph{INTEGRAL} operation in orbit. In addition to the variability maps, we have compiled significance maps in three energy bands (20-40, 40-100, and 100-200 keV). All our maps are available as an online service to the community using the SkyView engine. 

\section*{Acknowledgements}

Based on observations with INTEGRAL, an ESA project with instruments and science data centre funded by ESA member states (especially the PI countries: Denmark, France, Germany, Italy, Switzerland, Spain), Czech Republic and Poland, and with participation of Russia and the USA.

This work was supported by the Swiss National Science Foundation and the Swiss Agency for Development and Cooperation in the framework of the programme SCOPES - Scientific co-operation between Eastern Europe and Switzerland. The computational part of the work was done at VIRGO.UA\footnote{http://virgo.org.ua} and BITP\footnote{http://bitp.kiev.ua} computing resources.

We are grateful to the anonymous referee for the critical remarks which helped us improve the paper.

IT acknowledges the support from the INTAS YSF grant No.~06-1000014-6348. 

\section*{Appendix: Error Calulation on $V_r$}

We use the standard error propagation formula to find the error, $\sigma_{V_r}$, of the function $V_r = f(S_{int,s}^2=a, S_{int,b}^2=b, F_{s}=c, F_{b}=d)$ so :
\begin{equation}
\sigma_{V_r} = \sigma_{f} = \sqrt{\left(\frac{\partial f}{\partial a}\sigma_a\right)^2 + \left(\frac{\partial f}{\partial b}\sigma_b\right)^2 + \left(\frac{\partial f}{\partial c}\sigma_c\right)^2 + \left(\frac{\partial f}{\partial d}\sigma_d\right)^2},
\label{VERR}
\end{equation}
where $\sigma_a = \sigma_b = \Sigma$ and $\sigma_c = \sigma_d = \sigma$ for a given source. By substituting the appropriate values in Eq.~\ref{VERR} and by taking derivatives we find that:
\begin{equation}
\sigma_{V_r} = \sqrt{\frac{\Sigma^2}{2\left(F_s - F_b\right)^2 \left(S_{int,s}^2 - S_{int,b}^2 \right)} + \frac{2 \sigma^2 \left(S_{int,s}^2 - S_{int,b}^2 \right) }{\left(F_s - F_b\right)^4}}.
\end{equation}

\bibliographystyle{aa}
\bibliography{intvarcat}

\clearpage

\pagestyle{empty}
\setlength{\voffset}{+0.3in}
\setlength{\textwidth}{7.2in}

\begin{small}
\longtabL{1}{

\begin{landscape}
\setcounter{table}{2}
\begin{longtable}{llllllllllll}

\caption[short caption]{The catalog of variable sources detected by \emph{INTEGRAL}\label{varcat}} \\
\hline
\noalign{\smallskip}
\multicolumn{1}{l|}{Nr.} & \multicolumn{1}{l|}{Name} &\multicolumn{3}{c|}{20-40 keV} &\multicolumn{3}{c|}{40-100 keV} &\multicolumn{3}{c|}{100-200 keV} & Class \\
\cline{3-11}
\noalign{\smallskip}
\noalign{\smallskip}
\multicolumn{1}{l|}{} & \multicolumn{1}{l|}{} & V & F & \multicolumn{1}{l|}{E} & V & F & \multicolumn{1}{l|}{E} & V & F & \multicolumn{1}{l|}{E} &  \\
\multicolumn{1}{l|}{} & \multicolumn{1}{l|}{} &  & ($10^{-11}$ergs~s$^{-1}$~cm$^{-2}$) & \multicolumn{1}{l|}{(ksec)} &  & ($10^{-11}$ergs~s$^{-1}$~cm$^{-2}$) & \multicolumn{1}{l|}{(ksec)} &  & ($10^{-11}$ergs~s$^{-1}$~cm$^{-2}$) & \multicolumn{1}{l|}{(ksec)} &  \\
\hline
\hline
\noalign{\smallskip}
\noalign{\smallskip}
\endfirsthead
\caption{ continued.} \\
\hline
\noalign{\smallskip}
\multicolumn{1}{l|}{Nr.} & \multicolumn{1}{l|}{Name} &\multicolumn{3}{c|}{20-40 keV} &\multicolumn{3}{c|}{40-100 keV} &\multicolumn{3}{c|}{100-200 keV} & Class \\
\cline{3-11}
\noalign{\smallskip}
\noalign{\smallskip}
\multicolumn{1}{l|}{} & \multicolumn{1}{l|}{} & V & F & \multicolumn{1}{l|}{E} & V & F & \multicolumn{1}{l|}{E} & V & F & \multicolumn{1}{l|}{E} &  \\
\multicolumn{1}{l|}{} & \multicolumn{1}{l|}{} &  & ($10^{-11}$ergs~s$^{-1}$~cm$^{-2}$) & \multicolumn{1}{l|}{(ksec)} &  & ($10^{-11}$ergs~s$^{-1}$~cm$^{-2}$) & \multicolumn{1}{l|}{(ksec)} &  & ($10^{-11}$ergs~s$^{-1}$~cm$^{-2}$) & \multicolumn{1}{l|}{(ksec)} &  \\
\hline
\hline
\noalign{\smallskip}
\noalign{\smallskip}
\endhead
\endfoot

\input{paptabbodyg.tex}

\noalign{\smallskip}
\noalign{\smallskip}
\hline

\end{longtable}
\hspace{-6mm} The table gives the values of fractional variability of the sources, $V_r$, and the flux, $F$, in $10^{-11}$ergs~s$^{-1}$~cm$^{-2}$ and corresponding $2\sigma$ errors. The conversion factors\footnote{We used the OSA 7.0 ARFs and RMFs and simulated a source with a Crab-like spectrum. Then using XSPEC we found the conversion between count rate and flux. It should be noted that this method depends on the spectral shape of the source, and that there are several ARFs for different periods of the mission, so the fluxes cannot be fully reliable. However, we find this approach the most suitable for the current catalog paper.} are 1~c/s = $4.46 \times 10^{-11}$ergs~s$^{-1}$~cm$^{-2}$ (20-40 keV), 1~c/s = $9.247 \times 10^{-11}$ergs~s$^{-1}$~cm$^{-2}$ (40-100 keV), and 1~c/s = $3.638 \times 10^{-10}$ergs~s$^{-1}$~cm$^{-2}$ (100-200 keV). The exposure time of the sources, $E$, in ksec is given for each energy band only if the source was detected. The class column gives the source type according to the reference catalog. If the source is localized in more then one map then the weighted mean values of the flux and variability are given. The sources active during specific time periods and not detected at the total variability map are indicated with bold font. The sources detected only on the total variability map are indicated with italics. The superscript numbers show the time period map where the source is visible and asterix means that the source is not detected at the respective significance map. Lower limit for fractional variability is given in such a case. The superscript ``g'' indicates that the source was affected by the ``ghost''.

\end{landscape}

}

\end{small}
\clearpage

\end{document}

%% file: varnstab.tex
IGR J00245+6251 & GRB & $2.003\pm0.006$ & 1522.2 & 2 & 20-40 \\
IGR J00245+6251 & GRB & $1.455\pm0.016$ & 1771.9 & 2,3 & 40-100 \\
IGR J00245+6251 & GRB & $0.340\pm0.017$ & 1522.2 & 2 & 100-200 \\
RX J0137.7+5814 & UNID & $0.319\pm0.166$ & 2477.1 & T & 20-40 \\
IGR J08408-4503 & HMXB & $0.199\pm0.127$ & 2945.0 & T & 20-40 \\
IGR J11098-6457 & AGN & $0.408\pm0.096$ & 423.0 & 3 & 40-100 \\
IGR J11321-5311 & HMXB & $0.383\pm0.066$ & 870.5 & 2 & 40-100 \\
IGR J11321-5311 & HMXB & $0.226\pm0.051$ & 870.5 & 2 & 100-200 \\
HR 4492 & UNID & $0.571\pm0.045$ & 1012.5 & 1 & 20-40 \\
PKS 1241-399 & AGN & $0.405\pm0.274$ & 1312.7 & T & 20-40 \\
IGR J16248-4603 & UNID & $0.486\pm0.101$ & 917.8 & 3 & 40-100 \\

%% file: ghostab.tex
PSR B1509-58 & 0.338 & 0.048 & 0.011 &  12.0 & 1054.6 & ND & ND & ND & ND & ND & ND & ND & ND & ND & ND \\
IGR J16493-4348 & 1.392 & 0.382 & 0.011 &  27.0 & 2349.7 & ND & ND & ND & ND & ND & ND & ND & ND & ND & ND \\
4U 1543-624 & 1.049 & 0.425 & 0.013 &   9.0 & 716.7 & ND & ND & ND & ND & ND & ND & ND & ND & ND & ND \\
XTE J1807-294 & 3.492 & 1.143 & 0.032 & 159.0 & 4925.0 & ND & ND & ND & ND & ND & ND & ND & ND & ND & ND \\
IGR J17597-2201 & 0.608 & 0.049 & 0.020 & 108.0 & 5355.6 & 0.892 & 0.105 & 0.020 & 108.0 & 5355.6 & ND & ND & ND & ND & ND \\
IGR J16351-5806 & ND & ND & ND & ND & ND & $<2.720$ & 0.000 & 0.022 &  24.0 & 1098.9 & ND & ND & ND & ND & ND \\
4U 1705-32 & ND & ND & ND & ND & ND & 1.667 & 0.512 & 0.035 & 135.0 & 3839.1 & ND & ND & ND & ND & ND \\
XTE J1818-245 & 3.495 & 0.987 & 0.039 & 324.0 & 8391.2 & 6.571 & 0.000 & 0.076 & 324.0 & 4246.5 & ND & ND & ND & ND & ND \\
IGR J17464-3213 & 3.165 & 0.041 & 0.018 & 225.0 & 12428.2 & 2.427 & 0.058 & 0.018 & 225.0 & 12428.2 & 1.567 & 0.164 & 0.028 & 240.0 & 8521.1 \\

%% file: paptabbodyg.tex
1 & IGR J00245+6251 & $>27.579^{2*}$ & $<0.324$ & $1522.2$ & $>15.784^{2*,3*}$ & $<0.852$ & $1771.9$ & $>6.324^{2*}$ & $<1.957$ & $1522.2$ & GRB \\
2 & IGR J00291+5934 & $2.845\pm0.360^{2}$ & $2.369\pm0.212$ & $1681.1$ & $2.858\pm0.558^{2}$ & $3.345\pm0.462$ & $1681.1$ & $ND$ & $ND$ & $0.0$ & LMXB \\
3 & IGR J00370+6122 & $>4.019^{1*}$ & $<0.423$ & $812.0$ & $ND$ & $ND$ & $0.0$ & $ND$ & $ND$ & $0.0$ & HMXB \\
4 & SMC X-1 & $0.647\pm0.040^{1}$ & $26.196\pm1.155$ & $56.5$ & $0.641\pm0.339^{1}$ & $6.669\pm2.419$ & $56.5$ & $ND$ & $ND$ & $0.0$ & HMXB \\
5 & 3A 0114+650 & $1.404\pm0.064^{1,2,3}$ & $6.718\pm0.210$ & $2295.6$ & $1.542\pm0.215^{1,2,3}$ & $4.702\pm0.447$ & $2295.6$ & $ND$ & $ND$ & $0.0$ & HMXB \\
6 & H 0115+634 & $1.832\pm0.017^{1,3}$ & $48.370\pm0.318$ & $961.5$ & $1.890\pm0.107^{1,3}$ & $16.680\pm0.665$ & $961.5$ & $ND$ & $ND$ & $0.0$ & HMXB \\
7 & 4U 0142+614 & $1.383\pm0.858^{3}$ & $1.402\pm0.603$ & $245.1$ & $ND$ & $ND$ & $0.0$ & $ND$ & $ND$ & $0.0$ & HMXB \\
8 & RX J0146.9+6121 & $1.248\pm0.523^{1}$ & $1.749\pm0.501$ & $626.9$ & $ND$ & $ND$ & $0.0$ & $ND$ & $ND$ & $0.0$ & HMXB \\
9 & IGR J01583+6713 & $3.482\pm3.303^{2}$ & $0.565\pm0.378$ & $999.8$ & $ND$ & $ND$ & $0.0$ & $ND$ & $ND$ & $0.0$ & HMXB \\
10 & {\bf NGC 788} & $0.599\pm0.131^{3}$ & $2.996\pm0.398$ & $468.0$ & $ND$ & $ND$ & $0.0$ & $ND$ & $ND$ & $0.0$ & AGN \\
11 & SWIFT J0216.3+5128 & $>2.926^{2*}$ & $<0.801$ & $750.4$ & $ND$ & $ND$ & $0.0$ & $ND$ & $ND$ & $0.0$ & AGN \\
12 & {\bf NGC 1052} & $ND$ & $ND$ & $0.0$ & $>2.593^{1*}$ & $<1.265$ & $607.0$ & $ND$ & $ND$ & $0.0$ & AGN \\
13 & NGC 1365 & $1.435\pm0.866^{2}$ & $2.102\pm0.876$ & $188.9$ & $ND$ & $ND$ & $0.0$ & $ND$ & $ND$ & $0.0$ & AGN \\
14 & EXO 0331+530 & $0.741\pm0.004^{2}$ & $170.496\pm0.644$ & $333.0$ & $0.735\pm0.045^{2}$ & $31.100\pm1.332$ & $333.0$ & $ND$ & $ND$ & $0.0$ & HMXB \\
15 & X Per & $0.419\pm0.018^{1,3}$ & $19.009\pm0.437$ & $792.8$ & $0.420\pm0.028^{1,3}$ & $29.324\pm0.942$ & $792.8$ & $ND$ & $ND$ & $0.0$ & HMXB \\
16 & LMC X-4 & $1.388\pm0.075^{1,3}$ & $10.026\pm0.376$ & $412.9$ & $1.426\pm0.534^{1}$ & $3.075\pm0.808$ & $349.5$ & $ND$ & $ND$ & $0.0$ & HMXB \\
17 & Crab & $>0.209^{1,2,3}$ & $<645.163$ & $1591.8$ & $>0.208^{1,2,3}$ & $<780.058$ & $1591.8$ & $0.212\pm0.001^{1,2,3}$ & $529.148\pm1.576$ & $1591.8$ & SNR \\
18 & H 0614+091 & $0.428\pm0.045^{1,2,3}$ & $14.917\pm0.468$ & $2065.7$ & $0.668\pm0.116^{1,2}$ & $13.048\pm1.152$ & $974.1$ & $ND$ & $ND$ & $0.0$ & LMXB \\
19 & 3A 0656-072 & $1.292\pm0.130^{3}$ & $5.416\pm0.385$ & $560.3$ & $ND$ & $ND$ & $0.0$ & $ND$ & $ND$ & $0.0$ & HMXB \\
20 & EXO 0748-676 & $0.270\pm0.026^{1,3}$ & $14.279\pm0.595$ & $546.5$ & $0.262\pm0.062^{3}$ & $17.019\pm2.283$ & $68.8$ & $ND$ & $ND$ & $0.0$ & LMXB \\
21 & Ginga 0836-429 & $1.113\pm0.015^{1}$ & $23.026\pm0.218$ & $1360.8$ & $1.121\pm0.029^{1}$ & $24.152\pm0.437$ & $1360.8$ & $1.326\pm0.359^{1}$ & $6.269\pm1.160$ & $1360.8$ & LMXB \\
22 & Vela X-1 & $0.853\pm0.001^{1,2,3}$ & $153.607\pm0.167$ & $3413.2$ & $0.730\pm0.008^{1,2,3}$ & $46.951\pm0.338$ & $3413.2$ & $ND$ & $ND$ & $0.0$ & HMXB \\
23 & GRO J1008-57 & $2.569\pm0.341^{1,2}$ & $2.473\pm0.231$ & $2126.4$ & $2.230\pm0.935^{1}$ & $1.817\pm0.536$ & $1049.8$ & $ND$ & $ND$ & $0.0$ & HMXB \\
24 & {\bf IGR J10101-5654} & $>3.170^{2*}$ & $<0.762$ & $1096.9$ & $ND$ & $ND$ & $0.0$ & $ND$ & $ND$ & $0.0$ & HMXB \\
25 & 4U 1036-56 & $1.924\pm0.687^{1,3}$ & $0.938\pm0.224$ & $1372.9$ & $ND$ & $ND$ & $0.0$ & $ND$ & $ND$ & $0.0$ & HMXB \\
26 & Mrk 421 & $0.646\pm0.019^{2,3}$ & $14.443\pm0.290$ & $1190.2$ & $0.655\pm0.041^{2}$ & $17.456\pm0.773$ & $515.1$ & $ND$ & $ND$ & $0.0$ & AGN \\
27 & {\bf IGR J11098-6457} & $ND$ & $ND$ & $0.0$ & $>2.023^{3*}$ & $<1.865$ & $423.0$ & $ND$ & $ND$ & $0.0$ & AGN \\
28 & Cen X-3 & $0.994\pm0.008^{1,2,3}$ & $37.787\pm0.207$ & $2188.2$ & $1.007\pm0.147^{1,2,3}$ & $4.961\pm0.422$ & $2188.2$ & $ND$ & $ND$ & $0.0$ & HMXB \\
29 & IGR J11215-5952 & $>3.736^{1,2*}$ & $<0.613$ & $1774.7$ & $ND$ & $ND$ & $0.0$ & $ND$ & $ND$ & $0.0$ & HMXB \\
30 & IGR J11305-6256 & $>0.959^{1,2,3*}$ & $<2.205$ & $2193.3$ & $ND$ & $ND$ & $0.0$ & $ND$ & $ND$ & $0.0$ & HMXB \\
31 & {\bf IGR J11321-5311} & $ND$ & $ND$ & $0.0$ & $>3.119^{2*}$ & $<1.135$ & $870.5$ & $>2.651^{2*}$ & $<3.106$ & $870.5$ & HMXB \\
32 & HR 4492 & $>4.555^{1*}$ & $<0.559$ & $1012.5$ & $ND$ & $ND$ & $0.0$ & $ND$ & $ND$ & $0.0$ & UNID \\
33 & IGR J11435-6109 & $0.961\pm0.107^{2}$ & $4.467\pm0.348$ & $672.4$ & $1.041\pm0.312^{2}$ & $3.475\pm0.714$ & $672.4$ & $ND$ & $ND$ & $0.0$ & HMXB \\
34 & 1E 1145.1-6141 & $0.870\pm0.019^{1,2,3}$ & $14.913\pm0.222$ & $2367.0$ & $1.002\pm0.063^{1,2,3}$ & $10.579\pm0.447$ & $2367.0$ & $ND$ & $ND$ & $0.0$ & HMXB \\
35 & H 1145-619 & $1.723\pm0.335^{1}$ & $2.513\pm0.345$ & $1112.7$ & $ND$ & $ND$ & $0.0$ & $ND$ & $ND$ & $0.0$ & HMXB \\
36 & {\bf IGR J12042-0756} & $ND$ & $ND$ & $0.0$ & $>2.536^{3*}$ & $<2.663$ & $299.8$ & $ND$ & $ND$ & $0.0$ & AGN \\
37 & NGC 4151 & $0.738\pm0.036^{1,2,3}$ & $8.179\pm0.225$ & $1883.9$ & $0.662\pm0.053^{1,2,3}$ & $12.689\pm0.504$ & $1883.9$ & $ND$ & $ND$ & $0.0$ & AGN \\
38 & NGC 4388 & $0.330\pm0.044^{1,2}$ & $8.433\pm0.307$ & $1672.6$ & $0.342\pm0.074^{1,2}$ & $13.156\pm0.652$ & $1672.6$ & $ND$ & $ND$ & $0.0$ & AGN \\
39 & GX 301-2 & $0.949\pm0.003^{1,2,3}$ & $122.223\pm0.239$ & $2586.3$ & $0.676\pm0.031^{1,2,3}$ & $16.475\pm0.477$ & $2586.3$ & $ND$ & $ND$ & $0.0$ & HMXB \\
40 & 3C 273 & $0.297\pm0.029^{1,2,3}$ & $8.042\pm0.172$ & $2556.9$ & $0.361\pm0.050^{1,2,3}$ & $11.674\pm0.374$ & $2556.9$ & $ND$ & $ND$ & $0.0$ & AGN \\
41 & {\bf 1H 1249-637} & $>3.187^{3*}$ & $<0.674$ & $910.1$ & $ND$ & $ND$ & $0.0$ & $ND$ & $ND$ & $0.0$ & HMXB \\
42 & 3A 1246-588 & $0.416\pm0.082^{2}$ & $3.635\pm0.362$ & $924.5$ & $ND$ & $ND$ & $0.0$ & $ND$ & $ND$ & $0.0$ & LMXB \\
43 & {\bf 3C 279} & $ND$ & $ND$ & $0.0$ & $2.163\pm1.193^{2}$ & $1.508\pm0.572$ & $1434.5$ & $ND$ & $ND$ & $0.0$ & AGN \\
44 & NGC 4945 & $0.360\pm0.053^{1,2,3}$ & $7.998\pm0.263$ & $2821.4$ & $0.336\pm0.069^{1,2,3}$ & $14.424\pm0.547$ & $2821.4$ & $0.881\pm0.307^{1}$ & $10.756\pm2.390$ & $801.2$ & AGN \\
45 & {\bf IGR J13091+1137} & $>2.685^{1*}$ & $<1.281$ & $547.8$ & $ND$ & $ND$ & $0.0$ & $ND$ & $ND$ & $0.0$ & AGN \\
46 & {\bf NGC 5033} & $ND$ & $ND$ & $0.0$ & $>2.675^{1*}$ & $<1.320$ & $698.0$ & $ND$ & $ND$ & $0.0$ & AGN \\
47 & Cen A & $0.349\pm0.007^{1,2,3}$ & $23.707\pm0.241$ & $2330.7$ & $0.350\pm0.011^{1,2,3}$ & $35.375\pm0.516$ & $2330.7$ & $0.395\pm0.070^{1,3}$ & $29.876\pm1.567$ & $1757.8$ & AGN \\
48 & 4U 1323-62 & $0.585\pm0.047^{1,2,3}$ & $5.844\pm0.228$ & $2569.7$ & $0.765\pm0.153^{1,2,3}$ & $5.321\pm0.468$ & $2569.7$ & $ND$ & $ND$ & $0.0$ & LMXB \\
49 & {\bf MCG-06-30-015} & $1.579\pm0.738^{1}$ & $1.056\pm0.340$ & $663.7$ & $ND$ & $ND$ & $0.0$ & $>2.991^{1*}$ & $<2.992$ & $663.7$ & AGN \\
50 & IC 4329A & $0.521\pm0.056^{1,3}$ & $5.883\pm0.304$ & $1333.4$ & $0.766\pm0.147^{1}$ & $6.614\pm0.842$ & $646.9$ & $ND$ & $ND$ & $0.0$ & AGN \\
51 & {\bf IGR J14003-6326} & $>2.621^{2*}$ & $<0.779$ & $690.8$ & $ND$ & $ND$ & $0.0$ & $ND$ & $ND$ & $0.0$ & UNID \\
52 & Circinus Galaxy & $0.227\pm0.025^{1,3}$ & $9.493\pm0.272$ & $1516.8$ & $ND$ & $ND$ & $0.0$ & $ND$ & $ND$ & $0.0$ & AGN \\
53 & PSR B1509-58$^g$ & $0.338\pm0.048^{2}$ & $5.927\pm0.400$ & $1054.6$ & $ND$ & $ND$ & $0.0$ & $ND$ & $ND$ & $0.0$ & SNR \\
54 & Cir X-1 & $1.108\pm0.080^{1,2,3}$ & $4.474\pm0.207$ & $3758.3$ & $ND$ & $ND$ & $0.0$ & $ND$ & $ND$ & $0.0$ & LMXB \\
55 & H 1538-522 & $0.632\pm0.013^{1,2,3}$ & $14.329\pm0.185$ & $4749.1$ & $1.768\pm0.998^{3}$ & $3.039\pm1.173$ & $1069.5$ & $ND$ & $ND$ & $0.0$ & HMXB \\
56 & XTE J1543-568 & $>3.448^{1,3*}$ & $<0.718$ & $2474.4$ & $ND$ & $ND$ & $0.0$ & $ND$ & $ND$ & $0.0$ & HMXB \\
57 & {\bf 4U 1543-624$^g$} & $1.049\pm0.425^{2}$ & $1.797\pm0.485$ & $716.7$ & $ND$ & $ND$ & $0.0$ & $ND$ & $ND$ & $0.0$ & LMXB \\
58 & {\bf AX J1550.8-5418} & $>3.611^{1*,3}$ & $<0.534$ & $2725.9$ & $ND$ & $ND$ & $0.0$ & $ND$ & $ND$ & $0.0$ & LMXB \\
59 & XTE J1550-564 & $2.209\pm0.032^{1}$ & $29.752\pm0.308$ & $1511.0$ & $2.196\pm0.033^{1}$ & $56.789\pm0.611$ & $1511.0$ & $2.082\pm0.108^{1}$ & $44.222\pm1.618$ & $1511.0$ & LMXB \\
60 & H 1608-522 & $1.349\pm0.036^{1,2,3}$ & $9.703\pm0.180$ & $4193.8$ & $1.959\pm0.124^{1,2,3}$ & $8.120\pm0.358$ & $4193.8$ & $1.418\pm0.411^{2}$ & $7.190\pm1.430$ & $1659.8$ & LMXB \\
61 & IGR J16195-4945 & $1.451\pm0.447^{1,2,3}$ & $1.179\pm0.172$ & $4478.2$ & $ND$ & $ND$ & $0.0$ & $ND$ & $ND$ & $0.0$ & HMXB \\
62 & IGR J16194-2810 & $2.263\pm1.096^{1}$ & $1.401\pm0.467$ & $1214.1$ & $ND$ & $ND$ & $0.0$ & $ND$ & $ND$ & $0.0$ & LMXB \\
63 & Sco X-1 & $>0.265^{1,2,3}$ & $<449.583$ & $1627.5$ & $0.494\pm0.039^{1,2,3}$ & $16.011\pm0.519$ & $1627.5$ & $ND$ & $ND$ & $0.0$ & LMXB \\
64 & IGR J16207-5129 & $1.143\pm0.181^{1,2,3}$ & $2.206\pm0.177$ & $4166.5$ & $ND$ & $ND$ & $0.0$ & $ND$ & $ND$ & $0.0$ & HMXB \\
65 & {\bf IGR J16248-4603} & $ND$ & $ND$ & $0.0$ & $>3.039^{3*}$ & $<1.480$ & $917.8$ & $ND$ & $ND$ & $0.0$ & UNID \\
66 & SWIFT J1626.6-5156 & $>3.841^{2*}$ & $<0.391$ & $1614.9$ & $ND$ & $ND$ & $0.0$ & $ND$ & $ND$ & $0.0$ & UNID \\
67 & IGR J16283-4838 & $2.258\pm0.800^{2,3}$ & $0.961\pm0.222$ & $2520.1$ & $ND$ & $ND$ & $0.0$ & $ND$ & $ND$ & $0.0$ & HMXB \\
68 & IGR J16318-4848 & $1.073\pm0.014^{1,2,3}$ & $19.002\pm0.172$ & $4304.5$ & $1.030\pm0.044^{1,2,3}$ & $12.066\pm0.337$ & $4304.5$ & $ND$ & $ND$ & $0.0$ & HMXB \\
69 & IGR J16320-4751 & $0.836\pm0.020^{1,2,3}$ & $11.067\pm0.171$ & $4406.7$ & $0.845\pm0.169^{1,2,3}$ & $4.536\pm0.334$ & $4406.7$ & $ND$ & $ND$ & $0.0$ & HMXB \\
70 & 4U 1626-67 & $0.377\pm0.035^{2,3}$ & $11.789\pm0.500$ & $1016.9$ & $ND$ & $ND$ & $0.0$ & $ND$ & $ND$ & $0.0$ & LMXB \\
71 & {\bf IGR J16328-4726} & $2.269\pm0.389^{1,2}$ & $1.588\pm0.183$ & $3686.7$ & $2.506\pm0.955^{1}$ & $1.935\pm0.516$ & $1867.1$ & $ND$ & $ND$ & $0.0$ & UNID \\
72 & 4U 1630-47 & $1.244\pm0.014^{1}$ & $33.502\pm0.271$ & $1869.6$ & $0.925\pm0.020^{1}$ & $33.887\pm0.516$ & $1869.6$ & $0.812\pm0.099^{1}$ & $16.736\pm1.353$ & $1869.6$ & LMXB \\
73 & {\bf IGR J16351-5806$^g$} & $ND$ & $ND$ & $0.0$ & $>2.720^{2*}$ & $<1.109$ & $1098.9$ & $ND$ & $ND$ & $0.0$ & AGN \\
74 & IGR J16358-4726 & $3.978\pm1.084^{1}$ & $1.412\pm0.271$ & $1842.8$ & $ND$ & $ND$ & $0.0$ & $ND$ & $ND$ & $0.0$ & HMXB \\
75 & IGR J16393-4643 & $0.688\pm0.125^{1,2,3}$ & $3.622\pm0.172$ & $4594.9$ & $ND$ & $ND$ & $0.0$ & $ND$ & $ND$ & $0.0$ & HMXB \\
76 & H 1636-536 & $0.748\pm0.010^{1,2,3}$ & $20.582\pm0.196$ & $3418.8$ & $1.023\pm0.041^{1,2,3}$ & $14.417\pm0.387$ & $3418.8$ & $1.398\pm0.534^{1}$ & $5.944\pm1.532$ & $1433.9$ & LMXB \\
77 & IGR J16418-4532 & $1.107\pm0.122^{1,2,3}$ & $3.036\pm0.173$ & $5157.6$ & $ND$ & $ND$ & $0.0$ & $ND$ & $ND$ & $0.0$ & HMXB \\
78 & GX 340+0 & $0.554\pm0.007^{1,2,3}$ & $21.683\pm0.174$ & $5108.7$ & $ND$ & $ND$ & $0.0$ & $ND$ & $ND$ & $0.0$ & LMXB \\
79 & IGR J16465-4507 & $1.951\pm0.572^{1}$ & $1.371\pm0.276$ & $2088.5$ & $ND$ & $ND$ & $0.0$ & $>4.206^{1*}$ & $<2.032$ & $2088.5$ & HMXB \\
80 & IGR J16479-4514 & $1.879\pm0.166^{1,2,3}$ & $2.984\pm0.175$ & $5331.1$ & $2.421\pm0.629^{1,2,3}$ & $2.074\pm0.336$ & $5331.1$ & $ND$ & $ND$ & $0.0$ & HMXB \\
81 & IGR J16493-4348$^g$ & $1.392\pm0.382^{2}$ & $1.494\pm0.266$ & $2349.7$ & $ND$ & $ND$ & $0.0$ & $ND$ & $ND$ & $0.0$ & HMXB \\
82 & GRO J1655-40 & $1.771\pm0.037^{2}$ & $19.291\pm0.282$ & $3450.1$ & $1.828\pm0.046^{2}$ & $29.777\pm0.533$ & $3450.1$ & $1.836\pm0.172^{2}$ & $23.010\pm1.421$ & $3450.1$ & LMXB \\
83 & {\bf IGR J16558-4150} & $ND$ & $ND$ & $0.0$ & $>6.632^{1*}$ & $<0.744$ & $3418.6$ & $ND$ & $ND$ & $0.0$ & UNID \\
84 & Her X-1 & $0.847\pm0.008^{2,3}$ & $52.803\pm0.367$ & $474.7$ & $0.926\pm0.117^{2,3}$ & $9.388\pm0.812$ & $474.7$ & $ND$ & $ND$ & $0.0$ & LMXB \\
85 & OAO 1657-415 & $0.834\pm0.004^{1,2,3}$ & $48.793\pm0.174$ & $8861.0$ & $0.779\pm0.014^{1,2,3}$ & $31.754\pm0.323$ & $8861.0$ & $ND$ & $ND$ & $0.0$ & HMXB \\
86 & XTE J1701-462 & $1.167\pm0.088^{2,3}$ & $4.656\pm0.236$ & $3036.8$ & $ND$ & $ND$ & $0.0$ & $ND$ & $ND$ & $0.0$ & LMXB \\
87 & GX 339-4 & $2.029\pm0.019^{1,2,3}$ & $29.318\pm0.191$ & $3916.5$ & $2.061\pm0.029^{1,2,3}$ & $37.574\pm0.368$ & $3916.5$ & $1.888\pm0.136^{1,2,3}$ & $19.639\pm0.980$ & $3916.5$ & LMXB \\
88 & 4U 1700-377 & $1.037\pm0.002^{1,2,3}$ & $142.908\pm0.155$ & $11696.8$ & $1.045\pm0.004^{1,2,3}$ & $105.130\pm0.284$ & $11696.8$ & $1.236\pm0.134^{1,2,3}$ & $25.588\pm0.745$ & $11696.8$ & HMXB \\
89 & GX 349+2 & $0.445\pm0.026^{1,2,3}$ & $28.934\pm0.145$ & $11912.8$ & $ND$ & $ND$ & $0.0$ & $ND$ & $ND$ & $0.0$ & LMXB \\
90 & H 1702-429 & $1.038\pm0.025^{1,2,3}$ & $11.497\pm0.175$ & $7731.6$ & $1.481\pm0.087^{1,2,3}$ & $8.900\pm0.327$ & $7731.6$ & $ND$ & $ND$ & $0.0$ & LMXB \\
91 & {\bf 4U 1705-32$^g$} & $ND$ & $ND$ & $0.0$ & $1.667\pm0.512^{2}$ & $2.266\pm0.427$ & $3839.1$ & $ND$ & $ND$ & $0.0$ & LMXB \\
92 & {\bf IGR J17088-4008} & $3.498\pm2.326^{3}$ & $0.709\pm0.325$ & $2654.6$ & $ND$ & $ND$ & $0.0$ & $ND$ & $ND$ & $0.0$ & LMXB \\
93 & H 1705-440 & $1.018\pm0.020^{1,2,3}$ & $13.569\pm0.180$ & $6747.6$ & $2.001\pm0.164^{1,2,3}$ & $6.127\pm0.339$ & $6747.6$ & $ND$ & $ND$ & $0.0$ & LMXB \\
94 & IGR J17091-3624 & $ND$ & $ND$ & $0.0$ & $1.612\pm0.184^{1,2}$ & $4.985\pm0.290$ & $8825.8$ & $3.300\pm0.986^{1}$ & $5.201\pm0.981$ & $4820.5$ & LMXB \\
95 & XTE J1709-267 & $3.577\pm1.289^{1,3}$ & $0.674\pm0.151$ & $8605.2$ & $4.418\pm2.617^{3}$ & $1.157\pm0.472$ & $3714.3$ & $ND$ & $ND$ & $0.0$ & LMXB \\
96 & IGR J17098-3628 & $2.562\pm0.395^{2}$ & $2.814\pm0.239$ & $3993.2$ & $1.622\pm0.143^{1,2}$ & $5.875\pm0.289$ & $8820.5$ & $ND$ & $ND$ & $0.0$ & LMXB \\
97 & IGR J17195-4100 & $1.361\pm0.314^{1}$ & $1.812\pm0.253$ & $4014.1$ & $7.114\pm6.035^{1}$ & $0.761\pm0.452$ & $4014.1$ & $ND$ & $ND$ & $0.0$ & CV \\
98 & XTE J1720-318 & $2.654\pm0.361^{1}$ & $1.781\pm0.170$ & $4940.9$ & $2.601\pm0.436^{1}$ & $2.649\pm0.307$ & $4940.9$ & $ND$ & $ND$ & $0.0$ & LMXB \\
99 & IGR J17204-3554 & $ND$ & $ND$ & $0.0$ & $3.686\pm1.659^{1}$ & $1.129\pm0.347$ & $4845.8$ & $ND$ & $ND$ & $0.0$ & AGN \\
100 & IGR J17252-3616 & $1.162\pm0.040^{1,2,3}$ & $6.147\pm0.123$ & $12090.8$ & $ND$ & $ND$ & $0.0$ & $ND$ & $ND$ & $0.0$ & HMXB \\
101 & 4U 1722-30 & $0.383\pm0.013^{1,2,3}$ & $12.906\pm0.103$ & $12720.4$ & $0.507\pm0.046^{1,2,3}$ & $12.061\pm0.193$ & $12720.4$ & $ND$ & $ND$ & $0.0$ & LMXB \\
102 & {\bf IGR J17285-2922} & $ND$ & $ND$ & $0.0$ & $>6.885^{3*}$ & $<0.562$ & $3750.0$ & $ND$ & $ND$ & $0.0$ & UNID \\
103 & 3A 1728-169 & $0.445\pm0.075^{1,2,3}$ & $8.496\pm0.151$ & $12403.3$ & $ND$ & $ND$ & $0.0$ & $ND$ & $ND$ & $0.0$ & LMXB \\
104 & GX 354-0 & $0.800\pm0.004^{1,2,3}$ & $30.897\pm0.107$ & $12420.7$ & $1.450\pm0.030^{1,2,3}$ & $14.648\pm0.201$ & $12420.7$ & $ND$ & $ND$ & $0.0$ & LMXB \\
105 & GX 1+4 & $0.722\pm0.003^{1,2,3}$ & $36.923\pm0.105$ & $13270.5$ & $0.779\pm0.007^{1,2,3}$ & $36.042\pm0.198$ & $13270.5$ & $ND$ & $ND$ & $0.0$ & LMXB \\
106 & 4U 1730-335 & $1.824\pm0.101^{1,2}$ & $3.294\pm0.123$ & $8863.1$ & $1.784\pm0.402^{1}$ & $2.193\pm0.298$ & $4890.8$ & $ND$ & $ND$ & $0.0$ & LMXB \\
107 & SLX 1735-269 & $0.360\pm0.030^{1,2,3}$ & $7.334\pm0.095$ & $13037.9$ & $0.490\pm0.081^{1,2,3}$ & $7.536\pm0.180$ & $13037.9$ & $5.972\pm3.115^{3}$ & $2.667\pm0.972$ & $3757.6$ & LMXB \\
108 & 4U 1735-444 & $0.329\pm0.014^{1,2,3}$ & $17.596\pm0.213$ & $6450.1$ & $ND$ & $ND$ & $0.0$ & $ND$ & $ND$ & $0.0$ & LMXB \\
109 & IGR J17391-3021 & $6.000\pm1.000^{1,2,3}$ & $0.821\pm0.096$ & $12685.8$ & $>4.768^{1,3*}$ & $<0.735$ & $8667.8$ & $ND$ & $ND$ & $0.0$ & HMXB \\
110 & {\bf GRS 1736-297} & $3.057\pm1.767^{3}$ & $0.466\pm0.183$ & $3696.1$ & $ND$ & $ND$ & $0.0$ & $>8.061^{3*}$ & $<1.469$ & $3696.1$ & LMXB \\
111 & XTE J1739-285 & $1.638\pm0.164^{2}$ & $2.463\pm0.171$ & $4085.8$ & $1.695\pm0.353^{2}$ & $2.430\pm0.331$ & $4085.8$ & $ND$ & $ND$ & $0.0$ & LMXB \\
112 & IGR J17407-2808 & $1.908\pm0.703^{3}$ & $0.760\pm0.181$ & $3697.6$ & $ND$ & $ND$ & $0.0$ & $ND$ & $ND$ & $0.0$ & HMXB \\
113 & SLX 1737-282 & $0.698\pm0.105^{3}$ & $2.444\pm0.182$ & $3713.8$ & $ND$ & $ND$ & $0.0$ & $ND$ & $ND$ & $0.0$ & LMXB \\
114 & IGR J17419-2802 & $3.805\pm1.813^{2}$ & $0.507\pm0.168$ & $4118.9$ & $ND$ & $ND$ & $0.0$ & $ND$ & $ND$ & $0.0$ & UNID \\
115 & XTE J1743-363 & $1.346\pm0.151^{1,2}$ & $2.427\pm0.138$ & $8493.8$ & $ND$ & $ND$ & $0.0$ & $ND$ & $ND$ & $0.0$ & UNID \\
116 & 1E 1740.7-2942 & $0.885\pm0.007^{1,2,3}$ & $18.157\pm0.094$ & $12625.5$ & $0.911\pm0.009^{1,2,3}$ & $26.844\pm0.180$ & $12625.5$ & $0.895\pm0.117^{1,2,3}$ & $17.816\pm0.478$ & $12625.5$ & LMXB \\
117 & KS 1741-293 & $1.184\pm0.072^{1,2}$ & $3.122\pm0.110$ & $8973.9$ & $1.402\pm0.170^{1,2}$ & $3.289\pm0.209$ & $8973.9$ & $ND$ & $ND$ & $0.0$ & LMXB \\
118 & GRS 1741.9-2853 & $1.111\pm0.099^{2,3}$ & $2.542\pm0.124$ & $7708.5$ & $2.413\pm0.659^{3}$ & $1.906\pm0.357$ & $3664.7$ & $ND$ & $ND$ & $0.0$ & LMXB \\
119 & IGR J17453-2853 & $0.841\pm0.069^{3}$ & $3.474\pm0.181$ & $3684.8$ & $1.676\pm0.422^{3}$ & $2.183\pm0.356$ & $3684.8$ & $ND$ & $ND$ & $0.0$ & LMXB \\
120 & 1A 1742-294 & $0.787\pm0.014^{1,2,3}$ & $9.350\pm0.095$ & $12651.4$ & $1.240\pm0.069^{1,2,3}$ & $6.167\pm0.181$ & $12651.4$ & $ND$ & $ND$ & $0.0$ & LMXB \\
121 & IGR J17464-3213$^g$ & $3.165\pm0.041^{1,2,3}$ & $10.883\pm0.101$ & $12428.2$ & $2.427\pm0.058^{1,2,3}$ & $11.404\pm0.191$ & $12428.2$ & $1.567\pm0.164^{1,3}$ & $11.485\pm0.605$ & $8521.1$ & LMXB \\
122 & 1A 1743-288 & $1.470\pm0.104^{2}$ & $3.445\pm0.170$ & $4088.0$ & $1.675\pm0.297^{2}$ & $2.808\pm0.331$ & $4088.0$ & $ND$ & $ND$ & $0.0$ & LMXB \\
123 & IGR J17473-2721 & $5.072\pm0.348^{3}$ & $3.697\pm0.179$ & $3742.2$ & $5.230\pm0.632^{3}$ & $4.121\pm0.352$ & $3742.2$ & $ND$ & $ND$ & $0.0$ & LMXB \\
124 & SLX 1744-299 & $0.360\pm0.044^{1,2,3}$ & $6.000\pm0.094$ & $12718.3$ & $0.723\pm0.147^{1,2,3}$ & $4.641\pm0.180$ & $12718.3$ & $ND$ & $ND$ & $0.0$ & LMXB \\
125 & GX 3+1 & $0.346\pm0.030^{1,2,3}$ & $7.503\pm0.094$ & $13168.1$ & $ND$ & $ND$ & $0.0$ & $ND$ & $ND$ & $0.0$ & LMXB \\
126 & {\bf 1A 1744-361} & $>7.992^{1*}$ & $<0.278$ & $4585.0$ & $ND$ & $ND$ & $0.0$ & $ND$ & $ND$ & $0.0$ & LMXB \\
127 & AX J1749.1-2733 & $2.186\pm0.418^{1,2}$ & $0.967\pm0.110$ & $9206.1$ & $ND$ & $ND$ & $0.0$ & $ND$ & $ND$ & $0.0$ & HMXB \\
128 & GRO J1750-27 & $2.762\pm0.064^{3}$ & $10.984\pm0.181$ & $3774.3$ & $2.688\pm1.249^{3}$ & $1.139\pm0.356$ & $3774.3$ & $ND$ & $ND$ & $0.0$ & HMXB \\
129 & IGR J17497-2821 & $2.877\pm0.146^{3}$ & $5.008\pm0.180$ & $3687.3$ & $2.873\pm0.177^{3}$ & $8.093\pm0.352$ & $3687.3$ & $2.580\pm0.536^{3}$ & $6.894\pm0.961$ & $3687.3$ & LMXB \\
130 & SLX 1746-331 & $2.474\pm0.684^{1}$ & $0.891\pm0.159$ & $4824.6$ & $2.045\pm0.546^{1}$ & $1.746\pm0.293$ & $4824.6$ & $ND$ & $ND$ & $0.0$ & LMXB \\
131 & SAX J1750.8-2900 & $3.084\pm0.581^{3}$ & $1.377\pm0.183$ & $3676.5$ & $3.347\pm1.345^{3}$ & $1.283\pm0.358$ & $3676.5$ & $ND$ & $ND$ & $0.0$ & LMXB \\
132 & {\bf IGR J17507-2856} & $2.893\pm0.778^{3}$ & $0.972\pm0.183$ & $3655.3$ & $ND$ & $ND$ & $0.0$ & $ND$ & $ND$ & $0.0$ & UNID \\
133 & {\bf XTE J1751-305} & $>6.395^{2*}$ & $<0.263$ & $3990.4$ & $ND$ & $ND$ & $0.0$ & $ND$ & $ND$ & $0.0$ & LMXB \\
134 & SWIFT J1753.5-0127 & $0.558\pm0.004^{2,3}$ & $43.636\pm0.211$ & $3001.7$ & $0.587\pm0.005^{2,3}$ & $72.753\pm0.450$ & $3001.7$ & $0.974\pm0.052^{2}$ & $65.056\pm2.346$ & $1182.1$ & LMXB \\
135 & IGR J17544-2619 & $5.379\pm1.265^{1,2,3}$ & $0.586\pm0.095$ & $13107.9$ & $ND$ & $ND$ & $0.0$ & $ND$ & $ND$ & $0.0$ & HMXB \\
136 & IGR J17597-2201$^g$ & $0.608\pm0.049^{1}$ & $4.474\pm0.162$ & $5355.6$ & $0.892\pm0.105^{1}$ & $5.040\pm0.299$ & $5355.6$ & $ND$ & $ND$ & $0.0$ & LMXB \\
137 & GX 5-1 & $0.585\pm0.003^{1,2,3}$ & $33.376\pm0.097$ & $13357.4$ & $1.529\pm0.305^{1}$ & $2.415\pm0.278$ & $5134.8$ & $ND$ & $ND$ & $0.0$ & LMXB \\
138 & GRS 1758-258 & $0.333\pm0.002^{1,2,3}$ & $37.359\pm0.097$ & $13251.7$ & $0.363\pm0.003^{1,2,3}$ & $60.096\pm0.184$ & $13251.7$ & $0.430\pm0.058^{1,2,3}$ & $41.703\pm0.488$ & $13251.7$ & LMXB \\
139 & GX 9+1 & $0.298\pm0.025^{1,2,3}$ & $10.890\pm0.110$ & $14186.3$ & $ND$ & $ND$ & $0.0$ & $ND$ & $ND$ & $0.0$ & LMXB \\
140 & IGR J18027-2016 & $0.891\pm0.097^{1,2,3}$ & $3.316\pm0.111$ & $14175.0$ & $2.797\pm1.166^{3}$ & $1.447\pm0.404$ & $4427.4$ & $ND$ & $ND$ & $0.0$ & HMXB \\
141 & XTE J1807-294$^g$ & $3.492\pm1.143^{1}$ & $0.704\pm0.157$ & $4925.0$ & $ND$ & $ND$ & $0.0$ & $ND$ & $ND$ & $0.0$ & LMXB \\
142 & SGR 1806-20 & $2.026\pm0.219^{1,3}$ & $1.984\pm0.135$ & $9837.6$ & $1.677\pm0.222^{1}$ & $3.756\pm0.328$ & $5355.9$ & $ND$ & $ND$ & $0.0$ & GRB \\
143 & XTE J1810-189 & $2.825\pm0.280^{3}$ & $3.095\pm0.216$ & $4586.4$ & $2.980\pm0.562^{3}$ & $3.244\pm0.427$ & $4586.4$ & $ND$ & $ND$ & $0.0$ & HMXB \\
144 & SAX J1810.8-2609 & $1.768\pm0.089^{3}$ & $5.576\pm0.197$ & $3986.1$ & $1.866\pm0.173^{3}$ & $5.909\pm0.380$ & $3986.1$ & $ND$ & $ND$ & $0.0$ & LMXB \\
145 & GX 13+1 & $0.644\pm0.037^{1,2,3}$ & $7.601\pm0.134$ & $13306.4$ & $ND$ & $ND$ & $0.0$ & $ND$ & $ND$ & $0.0$ & LMXB \\
146 & 4U 1812-12 & $0.225\pm0.016^{1,2,3}$ & $18.049\pm0.146$ & $8364.3$ & $0.265\pm0.045^{1,2,3}$ & $22.116\pm0.293$ & $8364.3$ & $ND$ & $ND$ & $0.0$ & LMXB \\
147 & GX 17+2 & $0.502\pm0.003^{1,2,3}$ & $39.952\pm0.146$ & $10505.6$ & $ND$ & $ND$ & $0.0$ & $ND$ & $ND$ & $0.0$ & LMXB \\
148 & SWIFT J1816.7-1613 & $>4.363^{2*,3}$ & $<0.653$ & $8030.4$ & $ND$ & $ND$ & $0.0$ & $ND$ & $ND$ & $0.0$ & HMXB \\
149 & {\bf IGR J18173-2509} & $2.470\pm0.907^{1}$ & $0.732\pm0.174$ & $5089.1$ & $ND$ & $ND$ & $0.0$ & $ND$ & $ND$ & $0.0$ & CV \\
150 & XTE J1817-330 & $1.898\pm0.060^{2}$ & $9.320\pm0.209$ & $3665.6$ & $1.838\pm0.119^{2}$ & $8.828\pm0.401$ & $3665.6$ & $ND$ & $ND$ & $0.0$ & UNID \\
151 & XTE J1818-245$^g$ & $3.495\pm0.987^{2,3}$ & $0.803\pm0.150$ & $8391.2$ & $>6.571^{2*}$ & $<0.584$ & $4246.5$ & $ND$ & $ND$ & $0.0$ & UNID \\
152 & SAX J1818.6-1703 & $4.756\pm0.815^{1,2,3}$ & $1.161\pm0.135$ & $13305.8$ & $4.957\pm2.061^{1,3}$ & $1.147\pm0.308$ & $9381.5$ & $ND$ & $ND$ & $0.0$ & HMXB \\
153 & AX J1820.5-1434 & $2.838\pm0.743^{1,3}$ & $1.072\pm0.170$ & $7942.9$ & $ND$ & $ND$ & $0.0$ & $ND$ & $ND$ & $0.0$ & HMXB \\
154 & H 1820-303 & $0.317\pm0.009^{1,2,3}$ & $25.088\pm0.124$ & $12446.9$ & $2.029\pm0.416^{2}$ & $3.018\pm0.407$ & $3717.7$ & $ND$ & $ND$ & $0.0$ & LMXB \\
155 & {\bf H 1822-000} & $1.648\pm0.518^{1,3}$ & $1.163\pm0.188$ & $5190.6$ & $ND$ & $ND$ & $0.0$ & $ND$ & $ND$ & $0.0$ & LMXB \\
156 & 3A 1822-371 & $0.274\pm0.013^{1,2,3}$ & $21.952\pm0.161$ & $10735.5$ & $ND$ & $ND$ & $0.0$ & $ND$ & $ND$ & $0.0$ & LMXB \\
157 & IGR J18259-0706 & $2.909\pm1.485^{3}$ & $0.646\pm0.227$ & $2944.8$ & $ND$ & $ND$ & $0.0$ & $ND$ & $ND$ & $0.0$ & AGN \\
158 & Ginga 1826-24 & $0.227\pm0.002^{1,2,3}$ & $58.774\pm0.131$ & $13300.6$ & $0.241\pm0.006^{1,2,3}$ & $57.864\pm0.245$ & $13300.6$ & $ND$ & $ND$ & $0.0$ & LMXB \\
159 & IGR J18325-0756 & $1.419\pm0.233^{1,2}$ & $2.014\pm0.189$ & $4556.7$ & $ND$ & $ND$ & $0.0$ & $ND$ & $ND$ & $0.0$ & UNID \\
160 & XB 1832-330 & $0.381\pm0.049^{2}$ & $6.841\pm0.259$ & $3285.9$ & $ND$ & $ND$ & $0.0$ & $ND$ & $ND$ & $0.0$ & LMXB \\
161 & Ser X-1 & $0.323\pm0.049^{1,2,3}$ & $7.551\pm0.143$ & $7700.9$ & $ND$ & $ND$ & $0.0$ & $ND$ & $ND$ & $0.0$ & LMXB \\
162 & {\bf IGR J18406-0539} & $3.930\pm1.616^{1}$ & $0.881\pm0.254$ & $2746.4$ & $>5.663^{2*}$ & $<0.805$ & $2321.6$ & $ND$ & $ND$ & $0.0$ & HMXB \\
163 & IGR J18410-0535 & $3.862\pm1.187^{1,2}$ & $0.860\pm0.182$ & $5097.1$ & $>4.837^{1,2*}$ & $<0.885$ & $5097.1$ & $ND$ & $ND$ & $0.0$ & HMXB \\
164 & {\bf AX J1841.3-0455} & $ND$ & $ND$ & $0.0$ & $1.360\pm0.356^{2}$ & $3.408\pm0.531$ & $2398.7$ & $4.763\pm2.214^{2}$ & $4.626\pm1.457$ & $2398.7$ & HMXB \\
165 & IGR J18450-0435 & $2.009\pm0.576^{1,2,3}$ & $1.080\pm0.147$ & $8550.2$ & $ND$ & $ND$ & $0.0$ & $ND$ & $ND$ & $0.0$ & HMXB \\
166 & Ginga 1843+009 & $1.983\pm0.193^{1,2,3}$ & $2.215\pm0.137$ & $8557.6$ & $>2.347^{2,3*}$ & $<2.300$ & $5584.1$ & $ND$ & $ND$ & $0.0$ & HMXB \\
167 & IGR J18483-0311 & $2.092\pm0.155^{1,2,3}$ & $2.936\pm0.145$ & $8745.6$ & $2.019\pm0.536^{1,2,3}$ & $2.346\pm0.290$ & $8745.6$ & $ND$ & $ND$ & $0.0$ & HMXB \\
168 & 3A 1850-087 & $0.606\pm0.129^{1,3}$ & $3.514\pm0.201$ & $5609.1$ & $ND$ & $ND$ & $0.0$ & $ND$ & $ND$ & $0.0$ & LMXB \\
169 & IGR J18539+0727 & $>3.774^{1,2*}$ & $<0.541$ & $4659.1$ & $4.301\pm2.238^{1}$ & $1.011\pm0.365$ & $2549.1$ & $ND$ & $ND$ & $0.0$ & UNID \\
170 & V1223 Sgr & $0.531\pm0.088^{3}$ & $5.267\pm0.360$ & $3038.9$ & $ND$ & $ND$ & $0.0$ & $ND$ & $ND$ & $0.0$ & CV \\
171 & XTE J1855-026 & $0.915\pm0.031^{1,2,3}$ & $7.383\pm0.145$ & $8413.8$ & $1.007\pm0.184^{1,2,3}$ & $5.454\pm0.288$ & $8413.8$ & $ND$ & $ND$ & $0.0$ & HMXB \\
172 & XTE J1858+034 & $2.488\pm0.065^{1,2}$ & $7.716\pm0.142$ & $5092.5$ & $2.337\pm0.882^{1}$ & $1.435\pm0.359$ & $2739.9$ & $ND$ & $ND$ & $0.0$ & HMXB \\
173 & HETE J1900.1-2455 & $0.492\pm0.013^{2,3}$ & $19.234\pm0.284$ & $4331.0$ & $0.578\pm0.035^{2,3}$ & $21.219\pm0.566$ & $4331.0$ & $ND$ & $ND$ & $0.0$ & LMXB \\
174 & 4U 1901+03 & $1.090\pm0.008^{1}$ & $34.868\pm0.187$ & $2724.2$ & $1.199\pm0.144^{1}$ & $4.586\pm0.357$ & $2724.2$ & $ND$ & $ND$ & $0.0$ & HMXB \\
175 & {\bf XTE J1908+094} & $ND$ & $ND$ & $0.0$ & $2.471\pm0.797^{1}$ & $1.698\pm0.362$ & $2339.5$ & $ND$ & $ND$ & $0.0$ & LMXB \\
176 & H 1907+097 & $0.744\pm0.054^{1,2,3}$ & $10.273\pm0.124$ & $6332.8$ & $ND$ & $ND$ & $0.0$ & $ND$ & $ND$ & $0.0$ & HMXB \\
177 & 4U 1909+07 & $0.594\pm0.016^{1,2,3}$ & $9.089\pm0.121$ & $6595.4$ & $0.607\pm0.137^{1,2,3}$ & $6.522\pm0.242$ & $6595.4$ & $ND$ & $ND$ & $0.0$ & HMXB \\
178 & {\bf IGR J19112+1358} & $>4.516^{3*}$ & $<0.379$ & $2251.2$ & $ND$ & $ND$ & $0.0$ & $ND$ & $ND$ & $0.0$ & UNID \\
179 & Aql X-1 & $1.928\pm0.048^{1,2,3}$ & $7.825\pm0.137$ & $7327.5$ & $1.927\pm0.095^{1,2,3}$ & $8.334\pm0.272$ & $7327.5$ & $ND$ & $ND$ & $0.0$ & LMXB \\
180 & SS 433 & $0.729\pm0.033^{1,2,3}$ & $5.876\pm0.123$ & $6809.4$ & $0.866\pm0.171^{3}$ & $4.911\pm0.528$ & $2102.8$ & $ND$ & $ND$ & $0.0$ & HMXB \\
181 & IGR J19140+0951 & $1.630\pm0.082^{1,2,3}$ & $6.169\pm0.124$ & $6314.8$ & $1.734\pm0.178^{1,2,3}$ & $4.446\pm0.249$ & $6314.8$ & $ND$ & $ND$ & $0.0$ & HMXB \\
182 & GRS 1915+105 & $0.506\pm0.001^{1,2,3}$ & $182.501\pm0.128$ & $6278.2$ & $0.568\pm0.002^{1,2,3}$ & $100.967\pm0.256$ & $6278.2$ & $0.686\pm0.065^{2,3}$ & $34.835\pm0.945$ & $4015.8$ & LMXB \\
183 & 4U 1916-053 & $0.391\pm0.083^{1,2}$ & $5.909\pm0.221$ & $3788.5$ & $ND$ & $ND$ & $0.0$ & $ND$ & $ND$ & $0.0$ & LMXB \\
184 & KS 1947+300 & $1.317\pm0.091^{1,2}$ & $8.906\pm0.424$ & $1376.9$ & $1.201\pm0.221^{1,2}$ & $7.845\pm0.809$ & $1376.9$ & $ND$ & $ND$ & $0.0$ & HMXB \\
185 & 3A 1954+319 & $1.018\pm0.035^{2,3}$ & $12.188\pm0.287$ & $2146.0$ & $0.774\pm0.128^{3}$ & $7.215\pm0.692$ & $1498.8$ & $ND$ & $ND$ & $0.0$ & LMXB \\
186 & Cyg X-1 & $>0.265^{1,2,3}$ & $<505.064$ & $3255.1$ & $>0.289^{1,2,3}$ & $<742.342$ & $3255.1$ & $0.327\pm0.002^{1,2,3}$ & $505.242\pm1.283$ & $3255.1$ & HMXB \\
187 & EXO 2030+375 & $2.218\pm0.017^{1,2,3}$ & $36.696\pm0.203$ & $3116.2$ & $2.122\pm0.050^{1,2,3}$ & $24.136\pm0.399$ & $3116.2$ & $>5.735^{3*}$ & $<3.575$ & $1092.4$ & HMXB \\
188 & Cyg X-3 & $0.357\pm0.001^{1,2,3}$ & $110.404\pm0.197$ & $3130.7$ & $0.338\pm0.005^{1,2,3}$ & $56.916\pm0.390$ & $3130.7$ & $ND$ & $ND$ & $0.0$ & HMXB \\
189 & SAX J2103.5+4545 & $1.497\pm0.045^{1,3}$ & $9.366\pm0.198$ & $2927.3$ & $1.617\pm0.150^{1,3}$ & $6.412\pm0.406$ & $2927.3$ & $ND$ & $ND$ & $0.0$ & HMXB \\
190 & IGR J21247+5058 & $0.270\pm0.039^{3}$ & $5.227\pm0.242$ & $1564.6$ & $ND$ & $ND$ & $0.0$ & $ND$ & $ND$ & $0.0$ & AGN \\
191 & {\bf 4U 2129+12} & $ND$ & $ND$ & $0.0$ & $1.313\pm0.806^{2}$ & $3.870\pm1.645$ & $133.0$ & $ND$ & $ND$ & $0.0$ & LMXB \\
192 & SS Cyg & $1.005\pm0.205^{1}$ & $2.413\pm0.317$ & $1147.2$ & $ND$ & $ND$ & $0.0$ & $ND$ & $ND$ & $0.0$ & CV \\
193 & Cyg X-2 & $0.421\pm0.017^{1,2,3}$ & $17.769\pm0.265$ & $2793.1$ & $ND$ & $ND$ & $0.0$ & $ND$ & $ND$ & $0.0$ & LMXB \\
194 & 3A 2206+543 & $1.043\pm0.071^{1,2,3}$ & $5.679\pm0.224$ & $2923.0$ & $0.946\pm0.276^{1,2,3}$ & $5.365\pm0.485$ & $2923.0$ & $ND$ & $ND$ & $0.0$ & HMXB \\
195 & 3C 454.3 & $0.369\pm0.063^{2}$ & $6.229\pm0.572$ & $271.3$ & $ND$ & $ND$ & $0.0$ & $ND$ & $ND$ & $0.0$ & AGN \\
196 & {\it gam Cas} & $0.358\pm0.138$ & $2.981\pm0.177$ & $2741.3$ & $ND$ & $ND$ & $0.0$ & $ND$ & $ND$ & $0.0$ & HMXB \\
197 & {\it RX J0137.7+5814} & $>4.075^{1*}$ & $<0.349$ & $2477.1$ & $ND$ & $ND$ & $0.0$ & $ND$ & $ND$ & $0.0$ & UNID \\
198 & {\it Vela Pulsar} & $0.207\pm0.080$ & $4.560\pm0.152$ & $2955.5$ & $ND$ & $ND$ & $0.0$ & $ND$ & $ND$ & $0.0$ & SNR \\
199 & {\it IGR J08408-4503} & $>3.937^{1*}$ & $<0.225$ & $2945.0$ & $ND$ & $ND$ & $0.0$ & $ND$ & $ND$ & $0.0$ & HMXB \\
200 & {\it NGC 4051} & $1.113\pm0.548$ & $1.249\pm0.265$ & $1483.4$ & $ND$ & $ND$ & $0.0$ & $ND$ & $ND$ & $0.0$ & AGN \\
201 & {\it PKS 1241-399} & $>3.411^{1*}$ & $<0.530$ & $1312.7$ & $ND$ & $ND$ & $0.0$ & $ND$ & $ND$ & $0.0$ & AGN \\
202 & {\it SWIFT J1922.7-1716} & $3.266\pm2.396$ & $0.640\pm0.269$ & $2804.3$ & $ND$ & $ND$ & $0.0$ & $ND$ & $ND$ & $0.0$ & UNID \\